\documentclass[article]{elsarticle}

\usepackage{lineno,hyperref}

\usepackage{algorithm}
\usepackage{algpseudocode}
\usepackage[lofdepth,lotdepth]{subfig}
\usepackage{framed}
\usepackage[normalem]{ulem}
\usepackage{textcomp}
\usepackage{pifont}

\modulolinenumbers[5]

\journal{Journal of Computational Physics}

\usepackage{color}

\newcommand{\E}[1]{{\color{black}#1}}
\newcommand{\Es}[1]{}

\newcommand{\half}{\frac{1}{2}}
\newcommand{\dif}{{\rm d}}
\newcommand{\dvol}{{\rm d}^3{\bf r}}
\newcommand{\dsur}{{\rm d}{\bf s}}

\newcommand{\vJ}{{\bf J}}

\newcommand{\vE}{{\bf E}}
\newcommand{\vB}{{\bf B}}
\newcommand{\vA}{{\bf A}}
\newcommand{\vH}{{\bf H}}
\newcommand{\vT}{{\bf T}}
\newcommand{\rotT}{\nabla\times{\bf T}}
\newcommand{\rotDT}{\nabla\times\Delta{\bf T}}
\newcommand{\rotDTp}{\nabla' \times\Delta{\bf T}'}
\newcommand{\vr}{{\bf r}}
\newcommand{\vg}{{\bf g}}

\newcommand{\tensor}[1]{\overline{\overline{#1}}}

\newcommand{\paral}{{\mkern3mu\vphantom{\perp}\vrule depth 0pt\mkern2mu\vrule depth 0pt\mkern3mu}}

\newcommand{\dvoln}{{\rm d}^nr}
\newcommand{\fui}{f^{(u_i)}}
\newcommand{\fuia}{f^{ ( u_i^{(\alpha)} ) }}
\newcommand{\fuij}{f^{(u_iu_j)}}
\newcommand{\fuiajb}{f^{( u_i^{(\alpha)}u_j^{(\beta)} )}}
\newcommand{\fuijb}{f^{( u_iu_j^{(\beta)} )}}
\newcommand{\fuip}{f^{(u_i')}}
\newcommand{\fuiap}{f^{ ( {u_i'}^{(\alpha)} ) }}
\newcommand{\hui}{h^{(u_i)}}
\newcommand{\huia}{h^{ ( u_i^{(\alpha)} ) }}

\usepackage{tikz}
\usetikzlibrary{shapes,arrows,positioning,calc}

\tikzstyle{startstop} = [rectangle, rounded corners, minimum width=3cm, minimum height=1cm,text centered, draw=black, fill=red!30]
\tikzstyle{io} = [trapezium, trapezium left angle=70, trapezium right angle=110, minimum width=3cm, minimum height=1cm, text centered, draw=black, fill=blue!30]
\tikzstyle{process} = [rectangle, minimum width=3cm, minimum height=1cm, align=left, draw=black, fill=orange!30]
\tikzstyle{empty} = [rectangle, align=left]
\tikzstyle{decision} = [diamond, minimum width=3cm, minimum height=1cm, align=center, draw=black, fill=green!30]
\tikzstyle{arrow} = [thick,->,>=stealth]

\biboptions{numbers,sort&compress}

\begin{document}

\begin{frontmatter}

\title{3D computation of non-linear eddy currents: variational method and superconducting cubic bulk$^1$\fnref{legal}}
\fntext[legal]{Accepted version for publication at Journal of Computational Physics. Published version at \url{https://doi.org/10.1016/j.jcp.2017.05.001}. \textcopyright 2017. This manuscript version is made available under the CC-BY-NC-ND 4.0 license \url{http://creativecommons.org/licenses/by-nc-nd/4.0/}}

\author{E Pardo\fnref{myfootnote}}
\fntext[myfootnote]{Corresponding author.}
\ead{enric.pardo@savba.sk}

\author{M Kapolka}

\address{Institute of Electrical Engineering, Slovak Academy of Sciences,\\Dubravska 9, 84104 Bratislava, Slovakia}

\begin{abstract}
Computing the electric eddy currents in non-linear materials, such as superconductors, is \E{not straightforward}. The design of superconducting magnets and power applications needs electromagnetic computer modeling, being in many cases a three-dimensional (3D) problem. Since 3D problems require high computing times, novel time-efficient modeling tools are highly desirable. This article presents a novel computing modeling method based on a variational principle. The self-programmed implementation uses an original minimization method, which divides the sample into sectors. This speeds-up the computations with no loss of accuracy, while enabling efficient parallelization. This method could also be applied to model transients in linear materials or networks of non-linear electrical elements. As example, we analyze the magnetization currents of a cubic superconductor. This 3D situation remains unknown, in spite of the fact that it is often met in material characterization and bulk applications. We found that below the penetration field and in part of the sample, current flux lines are not rectangular and significantly bend in the direction parallel to the applied field. In conclusion, the presented numerical method is able to time-efficiently solve \E{fully} 3D situations \E{without loss of accuracy}.
\end{abstract}


\end{frontmatter}


\section{Introduction}
\label{s.intro}

Electrical eddy currents appear in conductors under varying magnetic fields, including the case of wires under alternating currents (AC) of sufficiently high frequency. In certain materials, such as superconductors, the resistivity is highly non-linear, an hence computing their response is \E{not straightforward} already in the quasi-magnetostatic situation \cite{acreview}.

Superconductors have been applied to magnet technology for decades and are promising for power applications, such as cables, fault-current limiters, transformers, generators, motors and levitations systems. An important issue of the design of these applications is the electromagnetic response under slowly changing magnetic fields or currents, usually for frequencies below 1 kHz. This design can only be done with computer modeling. In many cases, the situation of study is essentially a three dimensional (3D) problem \cite{acreview}, which involve time-extensive computations. Therefore, novel time-efficient 3D modeling tools are highly desirable.

Regarding material science, the magnetization currents in many situations is 3D, such as bulks shaped as rectanglar prisms, multi-granular samples, and multi-filamentary tapes with a conducting matrix. 3D modelling may also enlighten macroscopic flux cutting effects in the force-free configuration \cite{vlaskovlasov15PRB,mishev15SST}.

\E{There are several published 3D modelling results} for the finite-element method (FEM) in the following formulations: $\vH$ \cite{pecher04ICS,zehetmayer06SST,zhangM12SSTa,grilli13Cry,zermeno14SSTa,stenvall14SST,escamez16IES}, $\vA-\phi$ \cite{grilli05IES,lousberg09SST,fagnard16SST,campbell09SST,komi09PhC,farinon14SST}, $\vT-\Omega$ \cite{grilli05IES}, and $\vH$ with cohomology decomposition \cite{stenvall14SST}; being $\vH$ the magnetic field, $\vA$ and $\phi$ the vector and scalar potentials, and $\vT$ and $\Omega$ the current and magnetic potentials. All these approaches require solving the electromagnetic quantities at both the sample volume and surrounding air, setting boundary conditions far away from the sample. Then, only a portion of the degrees of freedom (DoF) are in the sample volume.

The DoF can be greatly reduced by methods taking the current density as state variable, since only the sample volume is taken into account. For mathematically 2D problems, this has been done by the variational method in $\bf J$ formulation \cite{prigozhin96JCP,prigozhin97IES,prigozhin98JCP,prigozhin11SST,HacIacinphase,pancaketheo,pardo15SST,sanchez06JAP,via15SST,ruuskanen14IES,zhangY15SST}, integral methods \cite{brandt95PRL,brandt95PRBa,brandt96PRB,rhyner98PhC,costa04SST,morandi15SST,vestgarden08PRB} and circuit methods \cite{vannugteren16IES}. The boundary-element/finite-element (BEM-FEM) method also avoids meshing the air \cite{russenschuck99rep,kurz02IES}. The FEM integral approach in the $\vT-\Omega$ formulation \E{has been} reduced to the sample region for 2D cross-sectional problems \cite{amemiya16SST} and 2D surfaces with 3D bending \cite{amemiya06JAP,nii12SST,ueda13IES}\E{.}\Es{; although this advantage is exclusive of thin films and cannot be extended to general 3D shapes.}

A possible variational method in 3D is very promising. The variational method in the Minimum Electro-Magnetic Entropy Production (MEMEP) implementation has been shown to be highly time efficient, presenting computing times scaling with only power 2 of the number of elements and being able to solve problems in 2D with up to half million DoF in the superconductor \cite{magnet10k}. Bossavit introduced the vartiational method in the $\vH$ formulation in 3D \cite{bossavit94IEM}, but did not solve any 3D example. \E{Elliott and Kashima provided further insight of the $\vH$ formulation, proposing a mixed formulation of magnetic field and magnetic potential \cite{elliott06JNA,kashima08MNA} and solved simple 3D examples.} Prigozhin developed the $\vJ$ formulation for 2D surfaces and cross-sectional problems \cite{prigozhin96JCP,prigozhin97IES,prigozhin98JCP}, which avoids taking DoF in the air. Badia and Lopez found that the functional minimizes the entropy production and introduced the Euler-Lagrange formalism \cite{badia01PRL,badia12SST}. Independently, Sanchez and Navau obtained a method to solve the Critical-State Model (CSM) in cylinders by minimization \E{of a certain magnetostatic energy} \cite{sanchez01PRB}. However, superconductors \E{in the CSM} only minimize the \E{magnetostatic} energy in \E{the initial curve from zero-field cool and} special situations \cite{tranarr,HacIacinphase}, being that method not applicable for arbitrarily non-uniform applied fields\E{, arbitrary cross-sections,} or simultaneous transport current and applied field, such as in a coil. \E{In any case, the involved mechanisms are irreversible.} A 3D variational principle in the $\vJ-\phi$ (or $\vJ-q$, where $q$ is the charge density) formalism was obtained in \cite{pardo15SST,couplingEUCAS}. Except for axi-symmetrical or infinitely long shapes, that method needs to compute $\vJ$ and $q$ iteratively, which increses the computing time \cite{couplingEUCAS}.

Independently on the numerical method, the magnetization currents in rectangular prisms of finite thickness remains mostly unknown, being a cube a particular case of this shape. Infinite rectangular prisms in the CSM were analytically solved in \cite{chen89JAP}. Thin rectangular films have been studied in \cite{brandt95PRL,brandt95PRBa} and \cite{prigozhin98JCP,navau08JAP} for an isotropic power-law $\vE(\vJ)$ relation and the CSM, respectively. Computations for a rectangular prism with a hole has been published in \cite{pecher04ICS} for a power-law $\vE(\vJ)$ relation. Reference \cite{badia05APL} presented approximated solutions for a cube in the CSM, assuming square current paths. The trapped field of an array of rectangular prisms is computed in \cite{zhangM12SSTa}. \E{Elliott and Kashima solved a rectangular prism \cite{elliott06JNA} and a sphere under rotating applied field \cite{kashima08MNA}, although these works practically do not discuss the results.}

This article presents a time-efficient 3D modeling tool based on a variational principle. This modeling tool for non-linear conductors is also efficient to compute transients in linear materials. It could also be easily adapted to modeling the response of networks of many non-linear electrical elements, such as diodes. As a computation example, we analyze a cubic bulk superconductor. We present the model in section \ref{s.mod}. Section \ref{s.varprin} details the deduction of a 3D variational principle in the $\vT$ formulation, which avoids spending DoF in the air and does not require solving the scalar potential or the charge density. The formalism also allows transport currents, in addition to the applied magnetic field. Although here we take an isotropic $\vE(\vJ)$ relation into account, the method also allows anisotropic $\vE(\vJ)$ relations, such as that for the force-free situation \cite{badia15SST}. Our self-programmed implementation uses a non-standard minimization method (section \ref{s.min}). This method has been greatly sped up with no loss of accuracy thanks to dividing the sample into sectors, which also enables efficient parallelization (section \ref{s.sectors}). The model is tested by comparing to analytical limits, showing good agreement (section \ref{s.tests}). Afterwards, we analyze the superconducting cube for both constant critical-current-density, $J_c$, (section \ref{s.constJc}) and magnetic-field-dependent $J_c$ (section \ref{s.JcB}). The appendices present details of variational calculus of functionals with double voulume integrals (\ref{s.delL}) and the discretization (\ref{s.vardis}).

Part of the results of this work have been presented in international conferences in 2015 and 2016 \cite{kapolka15EUCAS,pardo16HTSmod}, the mid-term report of M Kapolka PhD thesis \cite{kapolka16min} and benchmark 5 of the HTS modeling workgroup \cite{Modelling_website}.


\section{Model}
\label{s.mod}

In this section, we present the physical assumptions (section \ref{s.phmod}), the variational principle (section \ref{s.varprin}) and several aspects regarding the numerical method and implementation (sections \ref{s.dis}-\ref{s.mag}).


\subsection{Material properties and physical situation}
\label{s.phmod}

\begin{figure}[tbp]
\begin{center}
{\includegraphics[trim=47 0 50 5,clip,width=8.0 cm]{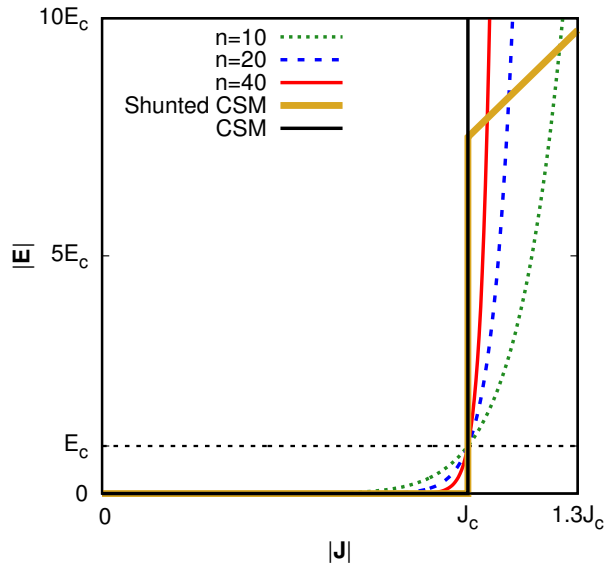}}
\caption{\label{f.EJ} \E{The isotropic power law ${\bf E}=E_c(|{\bf J|}/J_c)^n({\bf J}/|{\bf J}|)$ approaches to the critical-state model (CSM) $\vE(\vJ)$ relation for $n\to\infty$. The shunted CSM ${\bf E}({\bf J})$ relation in (\ref{shCSM}) models the case of a normal conducting material in parallel with the superconductor.}}
\end{center}
\end{figure}

Although the numerical method is valid for any vector $\vE(\vJ)$ relation of the material, \E{either isotropic or not}, in this work we consider an isotropic power law as
\begin{equation}
\vE(\vJ)=E_c \left ( \frac{|\vJ|}{J_c} \right )^n\frac{\vJ}{|\vJ|},
\label{EJ}
\end{equation}
where $E_c$ is an arbitrary constant, usually $10^{-4}$ V/m, $J_c$ is the critical current density, and $n$ is the power-law exponent. The limit of $n\to\infty$ corresponds to the isotropic critical-state model (CSM), which assumes a multi-valued $\vE(\vJ)$ relation, such as that of the CSM \E{(see figure \ref{f.EJ}). For the latter, 
\begin{equation}
\vE(\vJ)=
\left \{
\begin{array}{ll}
0\ \ & \textrm{if $|\vJ| < J_c$} \\
\infty & \textrm{if $|\vJ|> J_c$}
\end{array}
\right . 
\end{equation}
allowing any value of $|\vE|$ for $|\vJ|=J_c$ and being $\vE$ parallel to $\vJ$.
} 

In general, $J_c$ and $n$ in (\ref{EJ}) depend on the magnetic field\footnote{In this article, we do not take magnetic materials into account, and hence the magnetic field and magnetic flux density are proportional $\vH=\vB/\mu_0$ being $\mu_0$ the void permeability. In the text, we use ``magnetic field" to refer to both the magnetic field and magnetic flux density.} $\vB$. A typical magnetic-field dependence of $J_c$ for isotropic materials is Kim's \E{formula} \cite{kim62PRL}
\begin{equation}
J_c(\vB)=\frac{J_{c0}}{ \left ( 1+\frac{|\vB|}{B_0} \right )^m },
\end{equation}
where $J_{c0}$, $B_0$ and $m$ are constants.

\E{
Although in this work we focus on isotropic $\vE(\vJ)$ relations, the method is also suitable for $\vE(\vJ)$ relations with non-parallel $\vE$ and $\vJ$. There are two kinds of anisotropic $\vE(\vJ)$ relations. The simplest is a material with internal preferential directions of higher $J_c$, which can be characterized by a resistivity tensor $\tensor{\rho}_0(\vJ)$, so that 
\begin{equation}
\vE(\vJ)=\tensor{\rho}_0(\vJ)\vJ .
\end{equation}
Another situation is the $\vE(\vJ,\vB)$ relation that describes force-free effects, where $J_c$ is higher in the $\vB$ direction. For that case, we may use the $\vE(\vJ,\vB)$ relation proposed by Badia and Lopez \cite{badia15SST}
\begin{equation}
\vE(\vJ,\vB)=E_c \left [ { \frac{\vJ_\paral^2}{J_{c\paral}^2} + \frac{\vJ_\perp^2}{J_{c\perp}^2} } \right ]^{\frac{n-1}{2}} 
\left ( { \frac{J_{c\perp}}{J_{c\paral}}\frac{\vJ_\paral}{J_{c\paral}} + \frac{\vJ_\perp}{J_{c\perp}} } \right ) ,
\end{equation}
where $J_{c\paral}$ and $J_{c\perp}$ are $J_c$ in the directions parallel and perpendicular to $\vB$, respectively, $\vJ_\paral\equiv(\vJ\cdot{\bf e}_B){\bf e}_B$, $\vJ_\perp\equiv{\bf e}_B\times\vJ\times{\bf e}_B$, and ${\bf e}_B\equiv\vB/|\vB|$. A problem with this $\vE(\vJ,\vB)$ relation is that the parallel and perpendicular components of $\vJ$ are not well defined when $\vB=0$. Therefore, $\vE(\vJ,\vB)$ needs to be isotropic for $\vB=0$. A solution has been proposed by Kashima in \cite{kashima08MNA}, where an auxiliar isotropic term is introduced. Alternatively, we could set $J_{c\paral}(\vB)$ and $J_{c\perp}(\vB)$ dependences such that they are equal at $\vB=0$.
}

The computed examples in this article are for uniform applied magnetic fields, $\vB_a$; although the presented variational principle is also valid for transport currents. We consider that the applied field follows the $z$ direction \E{(figure \ref{film.fig}c)} and is generated by a long racetrack coil in the $y$ direction and high in the $z$ direction. The resulting applied vector potential $\vA_a$ in Coulomb's gauge, defined as Appendix B in \cite{acreview}, is
\begin{equation}
\vA_a(\vr)\approx B_ax{\bf e}_y,
\label{Aalong}
\end{equation}
where $B_a$ is such that $\vB_a=B_a{\bf e}_z$, and ${\bf e}_y,{\bf e}_z$ are the unit vectors in the $y$ and $z$ directions, respectively.

\subsection{Variational principle}
\label{s.varprin}	

In this section, we present a 3D variational principle where the computation is done in the superconductor (or conductor) volume only, excluding the surrounding air. Compared to previous functionals, we do not require the scalar potential or the charge density in order to find the current density \cite{pardo15SST,couplingEUCAS}. We also show that the functional always presents a minimum and that this minimum is unique. We name the variational principle and its 3D implementation as Minimum Electro-Magnetic Entropy Variation \E{in 3D} (MEMEP3D), since the solution minimizes the entropy production \cite{badia15SST,pardo15SST}.

Let consider that the material follows a certain non-linear vector $\vE(\vJ)$ relation. By now, we assume that $\vE(\vJ)$ is differentiable to the second order (sections \ref{s.Jphi} and \ref{s.T}). Later, we will show that the deduction is also valid if $\vE(\vJ)$ is multi-valued, such as in the CSM (section \ref{s.CSM}). The goal is to find a functional such that we can obtain $\vJ$ by minimizing that functional.

\subsubsection{$\vJ-\phi$ formulation}
\label{s.Jphi}

For a given electrostatic potential $\phi$ and Coulomb's gauge for the vector potential, the current density follows
\begin{equation}
\vE(\vJ)=-\frac{\partial\vA[\vJ]}{\partial t}-\frac{\partial\vA_a}{\partial t}-\nabla\phi , \label{Jeq}\\
\end{equation}
where $\vA[\vJ]$ is the vector potential in Coulomb's gauge created by $\vJ$, being
\begin{equation}
\vA[\vJ](\vr)=\frac{\mu_0}{4\pi}\int_V\dif V'\frac{\vJ(\vr')}{|\vr -\vr'|}.
\end{equation}
As shown in \cite{pardo15SST}, solving the equation above is the same as minimizing the following functional for the change of current density, $\Delta \vJ$, between two time steps, $t=t_0$ and $t=t_0+\Delta t$,
\begin{eqnarray}
L[\Delta\vJ] & = & \int_{V}\dif V\left ( \frac{1}{2}\Delta\vJ\cdot\frac{\vA[\Delta\vJ]}{\Delta t} \right . \nonumber\\
& + & \left . \Delta\vJ\cdot\frac{\Delta\vA_a}{\Delta t} + U(\vJ_0+\Delta\vJ)+\nabla\phi\cdot(\vJ_0+\Delta\vJ) \right ) \nonumber\\
& = & \int_V\dif V\int_V\dif V' \frac{\mu_0}{8\pi\Delta t}\frac{\Delta\vJ\cdot\Delta\vJ'}{|\vr-\vr'|} \nonumber \\
& + & \int_V\dif V \left ( \Delta\vJ\cdot\frac{\Delta\vA_a}{\Delta t} + U(\vJ_0+\Delta\vJ)+\nabla\phi\cdot(\vJ_0+\Delta\vJ) \right ),
\label{LJ}
\end{eqnarray}
where $V$ is the sample volume; $\dif V$ and $\dif V'$ are the volume differentials relative to $\vr$ and $\vr'$, respectively; $\vJ_0$ is the current density at time $t_0$; $\Delta\vJ'\equiv\Delta\vJ(\vr')$; $\vA[\Delta\vJ]$ is the vector potential created by $\Delta\vJ$ in Coulomb's gauge; and the dissipation factor $U(\vJ)$ is defined as
\begin{equation}
U(\vJ)\equiv\int_0^{\bf J}\dif \vJ'\cdot\vE(\vJ'),
\label{U}
\end{equation}
which is uniquely defined because $\nabla_{\vJ}\times{\vE}(\vJ)=0$ for any physical $\vE(\vJ)$ \cite{pardo15SST}. \E{For small $\Delta\vJ$, the dissipation factor is a measure of the energy dissipation due to $\Delta\vJ$, since $U(\vJ_0+\Delta\vJ)-U(\vJ_0)\approx \Delta\vJ\cdot\vE(\vJ_0)$.} For the power-law $\vE(\vJ)$ relation of (\ref{EJ}), the dissipation factor becomes
\begin{equation}
U(\vJ)=\frac{E_cJ_c}{n+1}\left ( \frac{|\vJ|}{J_c} \right )^{n+1}
.
\end{equation}

Next, we show that the physical $\Delta \vJ$ is an extreme of the functional (\ref{LJ}). The extreme occurs when 
the functional variation follows $\delta L[\Delta\vJ]=0$, where the variation is defined as (p. 192 of \cite{courant})
\begin{equation}
\delta L[\Delta{\bf J}]=\epsilon \left (  \frac{\dif}{\dif\epsilon} L[\Delta\vJ+\epsilon\vg ] \right )_{\epsilon=0},
\end{equation}
where $\epsilon$ is an arbitrary parameter with small value and $\vg(\vr)$ is any arbitrary function with continuous second derivatives except at the sample surface and vanishes outside the sample. Naturally, $\vg(\vr)$ should be non-zero at least at one point within the sample. Since functional (\ref{LJ}) contains a double volume integral, we cannot find the variation by applying the usual Euler equations but equations (\ref{EulerdV2}), deduced in the appendix. Thence, we obtain the variation
\begin{eqnarray}
\delta L[\Delta\vJ] & = & \epsilon \int_V\dif V \vg\cdot \int_V \dif V' \frac{\mu_0}{4\pi\Delta t} \frac{\Delta\vJ'}{|\vr-\vr'|} \nonumber \\
& + & \epsilon\int_V\dif V \vg\cdot \left ( { \frac{\Delta \vA_a}{\Delta t} + \vE(\vJ_0+\Delta\vJ) + \nabla\phi } \right ) \nonumber \\
& = & \epsilon \int_V\dif V\vg\cdot \left ( { \frac{\vA[\Delta\vJ]+\Delta \vA_a}{\Delta t} + \vE(\vJ_0+\Delta\vJ) + \nabla\phi } \right ).
\label{dLJ}
\end{eqnarray}
At the extreme of the functional $\delta L[\Delta\vJ]$=0 for any $\vg(\vr)$, and hence the expression within the paranthesis in (\ref{dLJ}) vanishes. This results in
\begin{equation}
\vE(\vJ_0+\Delta J)=-\frac{(\vA[\Delta\vJ]+\Delta\vA_a)}{\Delta t}-\nabla\phi,
\end{equation}
which is the time discretized form of equation (\ref{Jeq}).

In the following, we proof that the extreme is a minimum and that the minimum is unique. This is guaranteed if $\delta^2 L[\Delta\vJ]$ is always positive. From the definition of $\delta^2 L[\Delta\vJ]$, 
\begin{equation}
\delta^2 L\equiv \half \epsilon^2 \left ( \frac{\dif^2}{\dif\epsilon^2} L[\Delta\vJ+\epsilon\vg ] \right )_{\epsilon=0},
\end{equation}
we obtain a general formula for $\delta^2 L$ in \ref{s.delL}. Applying (\ref{d2Lboth}) to (\ref{LJ}) yields
\begin{eqnarray}
\delta^2L[\Delta\vJ] & = & \half\epsilon^2\int_V\dif V\int_V \dif V' \frac{\mu_0}{4\pi\Delta t}\frac{\vg(\vr)\cdot\vg(\vr')}{|\vr-\vr'|} \nonumber \\
& + & \half\epsilon^2\int_V\dif V\vg(\vr)\overline{\overline{\rho}}(\vJ_0+\Delta\vJ)\vg(\vr),
\label{d2LJ}
\end{eqnarray}
where $\overline{\overline{\rho}}$ is the differential resistivity tensor, with matrix elements $\rho_{ij}\equiv\partial E_i/\partial J_j$. From irreversible thermodynamic principles, the differential resistivity is positive \E{definite}, and hence $\vg\overline{\overline{\rho}}(\vJ_0+\Delta\vJ)\vg\ge 0$ for any $\vg$ and $\Delta\vJ$. The first term of (\ref{d2LJ}) is proportional to the magnetic self-interaction energy of a current density $\vg(\vr)$, and hence this term is always positive. As a consequence, $\delta^2L>0$ for any $\Delta\vJ$ and $\vg$. Then, the extreme of the functional is a minimum and it is unique. The uniqueness is due to the lack of maximums and saddle points, which are required for the existence of multiple minimums.

The main problem with this functional, equation (\ref{LJ}), is that the scalar potential, or $\nabla\phi$, should be known in order to obtain $\Delta{\bf J}$, with the exception of infinitely long or axi-symmetrical problems \cite{pardo15SST}. We may think to take the pair $(\Delta\vJ,\phi)$ as functions of $L$ in order to simultaneously obtain $\Delta\vJ$ and $\phi$. However, the functional does not present a minimum with respect to $\phi$. Although the variation with respect to $\phi$,
\begin{equation}
\delta L[\phi]=\epsilon \frac{\dif}{\dif\epsilon}L[\phi+\epsilon g]=\int_V\dif Vg \nabla\cdot(\vJ_0+\Delta\vJ),
\end{equation}
results in a physical Euler equation, 
\begin{equation}
\nabla\cdot\vJ=0, \label{divJ0}
\end{equation}
the second variation vanishes, $\delta^2L[\phi]=0$, for any $\phi$, and hence $\phi$ cannot be obtained as the minimizing $L$. 

\subsubsection{$\vT$ formulation}
\label{s.T}

A solution to decouple $\vJ$ and $\phi$ is the following. For samples subjected to an applied magnetic field only, without transport current, all current flux lines close within the conductor. Therefore, we can always consider $\vJ$ as magnetization currents from an effective magnetization $\vT$, such that \E{(see section 5.8 of \cite{jackson})}
\begin{equation}
\vJ=\nabla\times\vT.
\end{equation}
In this way, we ensure condition (\ref{divJ0}). Since we take $\vT$ as an effective magnetization, $\vT$ vanishes outside the sample. \E{At the surface, $\vT\times{\bf e}_n$ will represent an effective surface current density, being ${\bf e}_n$ the outward surface unit vector}. The taken physical model (section \ref{s.phmod}) assumes that there is no surface current density (only volume current density is present). Then, at the surface the parallel component of $\vT$ vanishes. We can take a transport current into account by taking a ``transport" contribution, $\vJ_t$,
\begin{equation}
\vJ=\nabla\times\vT+\vJ_t. \label{JTJt}
\end{equation}
Since the effective magnetization $\vT$ vanishes outside the sample, the net current $I$ is entirely due to $\vJ_t$
\begin{equation}
I=\int_S\dsur\cdot\vJ=\int_S\dsur\cdot\vJ_t,
\label{IJt}
\end{equation}
where $S$ is any surface that contains the cross-section of the conductor. In this way, we can find $\vJ$ by taking $\vJ_t$ as a given parameter that follows (\ref{IJt}) and $\nabla\cdot\vJ_t=0$, and afterwards find $\vT$ by minimizing the functional of (\ref{LJ}). For helical wires, for instance, $\vJ_t$ can be taken as uniform in the cross-section, following the spiral direction. \E{Then, the non-helical components of $\vJ$ \cite{stenvall13IES} are included in $\nabla\times\vT$, which is later solved by minimization.} For straight wires with variable cross-section or constrictions, $\vJ_t$ can be taken as homogeneous with a uniform cross-section and a large but finite resistance between the artificial homogeneous cross-section and the constriction; thus, the current redistribution in the constrictions is \E{again} contained within $\nabla\times\vT$. Once $\vT$ is found, $\nabla\phi$ can be found from equations (\ref{JTJt}) and (\ref{Jeq}). The functional (\ref{LJ}) with respect to the change in $\vT$ between two time steps, $\Delta\vT$, becomes
\begin{eqnarray}
L[\Delta\vT] & = & \int_V\dif V\left (  \half \rotDT\cdot\frac{\vA[\rotDT]}{\Delta t} + \rotDT\cdot\frac{(\Delta\vA_a+\Delta\vA_t)}{\Delta t} \right . \nonumber\\
& + & U(\vJ_0+\Delta\vJ_t+\rotDT)+\nabla\phi\cdot(\vJ_0+\Delta\vJ_t+\rotDT) \Bigg) ,
\label{LT}
\end{eqnarray}
where $\Delta\vJ_t$ is the variation of the transport current density between two time steps, $\Delta\vT$ is such that $\Delta\vJ=\rotDT+\Delta\vJ_t$,, and $\Delta\vA_t$ and $\vA[\rotDT]$ are the vector potential generated by $\Delta\vJ_t$ and $\rotDT$, respectively. In equation (\ref{LT}) we ignored the terms independent on $\Delta\vT$. By vector analysis, it can be seen that the last term is
\begin{eqnarray}
&& \int_V\dif V \nabla\phi\cdot(\vJ_0+\Delta\vJ_t+\rotDT)=\int_V\dif V \nabla\phi\cdot(\vJ_t+\rotT) \nonumber \\
&& =\int_{S_i}\dsur\cdot(\phi\vJ_t) + \int_{S_o}\dsur\cdot(\phi\vJ_t),
\label{Lphi}
\end{eqnarray}
where $S_i$ and $S_o$ are the wire cross-sections where the transport current gets in and out, respectively. Taking $S_i$ and $S_o$ as equipotentials, the integral becomes $\Delta\phi I$, where $\Delta \phi$ is the voltage drop. In any case, since the term in (\ref{Lphi}) does not depend on $\Delta\vT$, it does not influence the minimization process. Therefore, this term can be dropped from the functional. Taking this into account, the functional in (\ref{LT}) becomes\E{
\begin{framed}
\begin{eqnarray}
L[\Delta\vT] & = & \int_V\dif V \Bigg( \rotDT\cdot\frac{\vA[\rotDT]}{2\Delta t} \nonumber \\
& + & \rotDT\cdot\frac{(\Delta\vA_a+\Delta\vA_t)}{\Delta t} + U(\vJ_0+\Delta\vJ_t+\rotDT) \Bigg) \nonumber \\
& = & \int_V\dif V \Bigg( \rotDT\cdot\frac{(\Delta\vA_a+\Delta\vA_t)}{\Delta t} + U(\vJ_0+\Delta\vJ_t+\rotDT) \Bigg) \nonumber\\
& + & \int_V\dif V\int_V\dif V' \frac{\mu_0}{8\pi\Delta t} \frac{(\rotDT)\cdot(\rotDTp)}{|\vr-\vr'|},
\label{LTint}
\end{eqnarray}
\end{framed}
\noindent being this a central result of the article.} \E{In the equation above,} we expanded the integral in $\vA[\rotDT]$ at the second step, \E{$\nabla'\times$ is the curl in the $\vr'$ frame}, and $\Delta\vT' \equiv \Delta\vT(\vr')$. By vector analysis and taking into account that $\vT$ vanishes outside the sample, we obtain the following alternative formulation \E{
\begin{framed}
\begin{eqnarray}
L[\Delta\vT] & = & \int_V \dif V \Bigg( \Delta\vT\cdot\frac{\vB[\Delta\vT]}{2\Delta t} + \Delta\vT\cdot \frac{\Delta\vB_a+\Delta\vB_t}{\Delta t} \nonumber \\
& + & U(\vJ_0+\Delta\vJ_t+\rotDT) \Bigg) , 
\label{LTB}
\end{eqnarray}
\end{framed}
}\noindent where $\Delta\vB_t$ and $\vB[\Delta\vT]$ are the magnetic field generated by $\Delta\vJ_t$ and $\rotDT$, being the latter defined as
\begin{eqnarray}
\vB[\Delta\vT] & = & \frac{\mu_0}{4\pi}\int_V\dif V' \frac{(\rotDTp)\times(\vr-\vr')}{|\vr-\vr'|^3} \nonumber \\
& = & \frac{\mu_0}{4\pi}\int_V\dif V' \frac{3{\bf n}({\bf n}\cdot\Delta\vT')-\Delta\vT'}{|\vr-\vr'|^3} 
\end{eqnarray}
where ${\bf n}\equiv(\vr-\vr')/|\vr-\vr'|$. Minimizing this functional in any of its fomulations, (\ref{LTint}) or (\ref{LTB}), corresponds to solving the Euler partial differential equation of this functional, corresponding to $\delta L[\Delta\vT]=0$. Applying equation (\ref{dLboth}) to (\ref{LTint}), we obtain
\begin{eqnarray}
\delta L[\Delta\vT] & = & \int_V\dif V \vg\cdot \int_V \dif V' \left ( \frac{\mu_0}{4\pi\Delta t}\frac{(\rotDTp)\times(\vr-\vr')}{|\vr-\vr'|^3} \right ) \nonumber \\
& & + \int_V\dif V \vg\cdot\left ( \frac{\Delta\vB_a+\Delta\vB_t}{\Delta t} + \nabla\times\vE \right ) \nonumber \\
& = & \int_V\dif V \vg \cdot \left ( \frac{ \vB[\Delta\vT]+\Delta\vB_a+\Delta\vB_t }{\Delta t} + \nabla\times\vE \right ) .
\label{dLT}
\end{eqnarray}
The extremal condition $\delta L[\Delta\vT]=0$ is fulfilled for any $\vg(\vr)$, if and only if
\begin{equation}
\nabla\times\vE(\vJ_0+\Delta\vJ_t+\rotDT)=-\frac{\vB[\Delta\vT]+\Delta\vB_a+\Delta\vB_t}{\Delta t} .
\end{equation}
Taking into account that the change of magnetic field is $\Delta\vB=\vB[\Delta\vT]+\Delta\vB_a+\Delta\vB_t$, the equation above is the discretized form of Faraday's law
\begin{equation}
\nabla\times\vE(\vJ_t+\rotT)=-\frac{\partial\vB[\vT]}{\partial t}-\frac{\partial\vB_a}{\partial t}-\frac{\partial\vB_t}{\partial t} .
\end{equation}
For this new functional, we can check again that the extreme is a minimum and it is unique by analyzing $\delta^2L[\Delta\vT]$. By applying equation (\ref{d2Lboth}) to (\ref{dLT}), we obtain the same $\delta^2L[\Delta\vJ]$ as in (\ref{d2LJ}) but replacing $\Delta\vJ$ by $\rotDT$ and $\vg$ by $\nabla\times\vg$. Following the same arguments as for (\ref{d2LJ}), $\delta^2L[\Delta\vT]>0$ for any $\Delta\vT$ and $\vg$, and hence the extreme is a minimum and it is unique. If instead of the functional in (\ref{LTint}) we use that in (\ref{LTB}), we obtain the same differential equation and the same conclusion regarding $\delta^2L[\Delta\vT]$.

\subsubsection{Critical-state model or $J_c(\vB)$ situations}
\label{s.CSM}

Although this reasoning assumes that the $\vE(\vJ)$ relation is differentiable up to second order, we can also apply the deduction above to the CSM. The reason is that we can approximate the CSM by the continuous $\vE(\vJ)$ relation of (\ref{EJ}), the limit of $n\to\infty$ corresponding to the CSM \E{(figure \ref{f.EJ})}. Since the deduction is valid for any $n$, whatever large, it will also be valid for the CSM. For the CSM, $U(\vJ)=0$ for $|\vJ|\le J_c$ and $U(\vJ)\to \infty$ for $|\vJ|>J_c$. In practice, one can solve the CSM by either setting $|\vJ|\le J_c$ as a constrain or taking 
\begin{equation}
U(\vJ)=
\left \{
\begin{array}{ll}
0\ \ & \textrm{if $|\vJ| \le J_c$} \\
\rho(|\vJ|^2-J_c^2)/2 & \textrm{if $|\vJ|\ge J_c$}
\end{array}
\right . 
\end{equation}
with a very large $\rho$, which has the physical interpretation of the normal-state resistivity \cite{bossavit94IEM}. \E{This dissipation function corresponds to the shunted CSM, with $\vE(\vJ)$ relation
\begin{equation}
\vE(\vJ)=
\left \{
\begin{array}{ll}
0\ \ & \textrm{if $|\vJ| < J_c$} \\
\rho\vJ & \textrm{if $|\vJ|> J_c$}
\end{array}
\right . ,
\label{shCSM}
\end{equation}
allowing any value of $|\vE|$ between 0 and $\rho|\vJ|$ for $|\vJ|=J_c$ (see figure \ref{f.EJ}).
}

The variational method above assumes that $\vE$ depends on $\vJ$ directly but it does not depend on $\vB$. For $\vE(\vJ,\vB)$, such as the power-law in (\ref{EJ}) with $J_c(\vB)$, $\vJ$ is found iteratively, as detailed in section \ref{s.min} and \ref{s.sectors}.



\subsection{Discretization}
\label{s.dis}

In this article, we minimize the functional in the form of equation (\ref{LTint}). We choose this option because this formalism may be more convenient for future situations with transport current. However, in this work we do not take transport currents into account, so that $\vJ_t=0$ in the formalism above.

\begin{figure}[ptb]
\centering
{\includegraphics[trim=0 0 0 0,clip,width=7.0 cm]{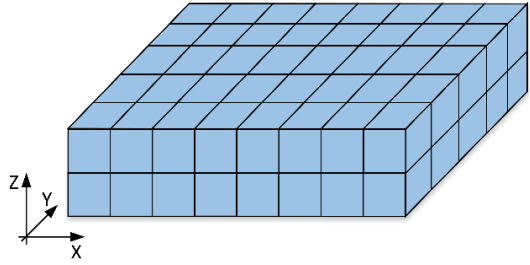}}
\caption{Dividing the sample uniformly into rectangular prisms greatly reduces the size of the interaction matrices.}
\label{discretization.fig}
\end{figure}

\begin{figure}[ptb]
\centering
 \subfloat[][]
{\includegraphics[trim=0 0 0 0,clip,height=5.0 cm]{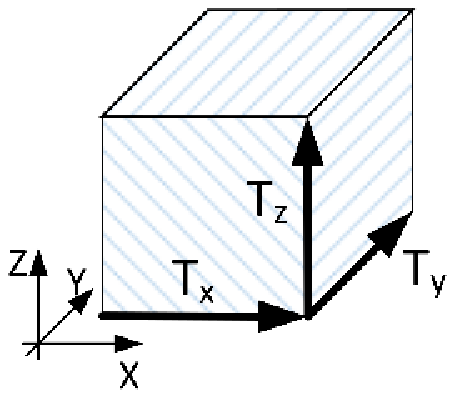}}
 \subfloat[][]
{\includegraphics[trim=-10 0 0 0,clip,height=5.5 cm]{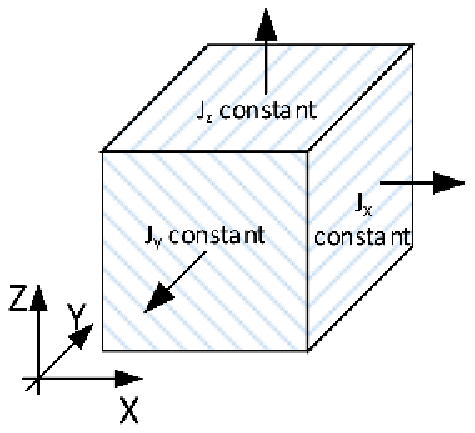}}
\caption{(a) The components of the effective magnetization $\vT$ are assumed uniform at the corresponding edges of the elements. (b) As a consequence, the components of the current density $\vJ$ are uniform at the corresponding surfaces of the elements.}
\label{Tvector.fig}
\end{figure}

The numerical method divides the sample into cells shaped as rectangular prisms (figure \ref{discretization.fig}). In this work, we use uniform mesh for all samples. Each cell contains edges and surfaces. The ${\bf{T}}$ vector, with components ${(T_{x},T_{y},T_{z})}$, is stored at the edges so that ${T_{x}}$ is saved at the edges parallel to the ${x}$ axis, and so on with ${T_{y}}$ and ${T_{z}}$ (figure \ref{Tvector.fig}a). We assume that ${\bf{T}}$ is constant along the length of the edge. The components of the current density, ${(J_{x},J_{y},J_{z})}$, are stored at the surfaces. Each cell surface contains the perpendicular component of the current density to the surface, which is assumed to be constant there (figure \ref{Tvector.fig}b). We can calculate ${\bf{T}}$ or ${\bf{J}}$ anywhere inside the cell by bi-linear and linear interpolation, respectively.

The components of the vector potential, $(A_x,A_y,A_z)$, are stored at the surfaces in the same way as for $\vJ$. The dissipation function $U$ in (\ref{U}) is assumed uniform at the cells volume. For a magnetic-field dependent critical current density, this dissipation factor depends on $\vB$, in addition to $\vJ$. For a consistent evaluation of $U$, we assume that $\vB$ is uniform at the cells.

For thin films, we take only one cell in the sample thickness. This results in averaging all electromagnetic quantities over the thickness, although the current still flows within the sample volume. Since at the surface the parallel component of $\vT$ vanishes, $\vT$ has only $z$ component.

More details on how to evaluate the relevant quantities from $\vT$ in this discretization, such as $\vJ$, $\vA$ and the functional are included in \ref{s.vardis}. As explained there, using uniform mesh allows to drastically reduce the size of the interaction matrices, minimizing computer memory requirements.

\subsection{Basic minimization method}
\label{s.min}

After increasing the time by $\Delta t$, the applied vector potential changes by $\Delta \vA_a$ (or the transport current density increases by $\Delta\vJ_t$, if present), which causes a change in $\vT$, $\Delta\vT$. This $\Delta\vT$ is found by minimizing functional (\ref{LTint}). At $t=0$, we consider the zero-field cool situation, and hence $\vA_a=\vJ_t=\vT=0$. The time increase does not need to be the same for all time steps \E{(time evolution in algorithm \ref{a.evolution})}.

\E{The basic minimization method is in algorithm \ref{a.basic}}.

First, we consider the change in $L[\Delta\vT]$ due to a change $v$ in the $s$ component of $\Delta\vT$ at edge $i$, $\Delta T_{si}$. By considering a positive $v$, this change of the functional, $dL_{si+}$, is evaluated at all edges. Afterwards, we also make the same evaluation but for a negative change, $-|v|$, resulting in a functional change $dL_{si-}$. Next, the algorithm chooses the edge where adding or removing $|v|$ to $\Delta T$ decreases the most the functional and sets the change in $\Delta T$ there. The process is repeated until changing $\Delta T$ at any edge increases the functional instead of decreasing it. 

The change in the functional, $dL_{si}$, depends only on the self-interaction energy of each edge, the vector potential at the intersecting surfaces to the edge omitting the change in $\Delta T_{si}$, and the modified $U$ at the neighboring cells (see \ref{s.vardis}). After finding the minimal edge and setting the change in $\Delta T_{si}$; $\Delta\vJ$, $U$ and $\Delta\vA$ are updated at the neighboring surfaces, neighboring cells, and all surfaces, respectively. This greatly accelerates the evaluation of $dL_{si}$ at the following steps because these quantities do not need to be calculated at each evaluation of $dL_{si}$, enabling computing complexity of only second order in the number of cells.

For a magnetic-field dependent $J_c$, the program finds a self-consistent solution by iteration \E{(not shown in the algorithms)}. After finding $\Delta\vT$ and its corresponding $\Delta\vJ=\rotDT$, the magnetic field is evaluated at the cells. Next, $\Delta\vT$ is found again by minimization and the process is repeated until the change in $\Delta \vT$ between two iterations is below a certain tolerance. In order to avoid oscillations, we apply a damping factor in the change of $\Delta\vT$ after each iteration.

When the routine converges for a given change in $\Delta T$, $|v|$, this $|v|$ is divided by 10 and the whole process is repeated again \E{(algorithm \ref{a.evolution})}. This reduction in $|v|$ is repeated until this value is below a certain pre-set tolerance. In this way, setting a tolerance 10 times stricter requires only twice (or less) computing time, achieving logarithmic complexity with respect to the tolerance. 

\begin{algorithm}
\caption{\E{The time, $t$, evolution algorithm allows any time division, which is determined by the function ``${\rm time}(k)$" being $k$ the time step. The slanted statement in blue could be either the basic minimization method of algorithm \ref{a.basic} or the iterative parallel routine in figure \ref{f.parallel}.}}\label{a.evolution}
\begin{algorithmic}[l]

\State \textbackslash * Time evolution algorithm * \textbackslash
\State $T_{si}:=0$ for all edges $s,i$;
\State Initialize $\vJ_t$ and $\vJ$ at all surfaces;
\State $t:=0$;
\For{$k=1$ to $n_{\rm time}$}
	\State $T_{0si}:=T_{si}$ at all edges $s,i$;
	\State $J_{0sj}:=J_{sj}$ at all surfaces $s,j$;
	\State $t_{\rm ini}:=t$;
	\State $t:=\rm{time}(k)$ ;
	\State $\Delta t:=t-t_{\rm ini}$;
  \State Update $\Delta\vJ_t$ at all surfaces;
	\State Update $\Delta\vA_a$ and $\Delta\vA_t$ at all edges;
	\State $\Delta T_{si}:=0$ for all edges $s,i$;
	\State $v:=10v_{\rm initial}$;
	\Repeat
		\State $v:=v/10$;
		\State {\color{blue} \sl Find $\Delta T_{si}$ for all $s$ and $i$ by minimization}
		\State {\color{blue} \sl and given minimum change $v$;}
	\Until{$v\le v_{\rm final}$}
	\State $T_{si}:=T_{0si}+\Delta T_{si}$ for all edges $s,i$;
	\State Update $\vJ$ at all surfaces;
	\State Post-process;
\EndFor

\end{algorithmic}
\end{algorithm}

\begin{algorithm}
\caption{\E{The basic minimization routine to find the change in $\vT$ between two time steps, $\Delta T_{si}$, at all edges $i$ of type $s$ scales only as the square of the total number of edges. $^*$ For the iterative parallel routine, the update of $\Delta\vA$ is done only in the surfaces of a single sector.}}\label{a.basic}
\begin{algorithmic}[l]

\State \textbackslash * Basic minimization method to obtain $\Delta T_{si}$ at all edges $i$ of type $s$ * \textbackslash
\Repeat
	\State $dL:=1$;
	\For{$s=1$ to 3}
		\State Find edge $i_+$ of type $s$ where adding $|v|$ to $\Delta T_{si}$
		\State produces the smallest $dL_{si+}$;
		\State Find edge $i_-$ of type $s$ where adding $-|v|$ to $\Delta T_{si}$
		\State produces the smallest $dL_{si-}$;
		\If{$dL_{si+}< dL_{si-}$}
			\State $dL_s:=dL_{si+}$;  $h_s:=|v|$;  $i_s:=i_+$;
		\Else
			\State $dL_s:=dL_{si-}$;  $h_s:=-|v|$;  $i_s:=i_-$;
		\EndIf
		\If{$dL_s<dL$}
			\State $dL:=dL_s$; $h_{\rm min}:=h_s$; $s_{\rm min}:=s$; $i_{\rm min}:=i_s$;
		\EndIf
	\EndFor
	\If{$dL<0$}
		\State $\Delta T_{s_{\rm min}i_{\rm min}} := \Delta T_{s_{\rm min}i_{\rm min}} + h_{\rm min}$;
		\State Update $\Delta\vJ$ at neighbouring surfaces of edge $i_{\rm min}$ of type $s_{\rm min}$;
		\State Update $U$ at neighbouring cells of edge $i_{\rm min}$ of type $s_{\rm min}$;
		\State Update $\Delta\vA$ at all surfaces$^*$;
	\EndIf
\Until{$dL\ge 0$}

\end{algorithmic}
\end{algorithm}


\subsection{Parallel minimization by sectors}
\label{s.sectors}

\begin{figure}[ptb]
\begin{tikzpicture}[node distance = 2cm, auto]
	\node (start) [startstop] {Start};
	\node (Jinit) [process, below of=start, node distance=1.5 cm] {Initial $\Delta T_{si}$};
	\node (Jp) [process, below of=Jinit,node distance=1.5 cm] {$\Delta T_{{\rm p},si}:=\Delta T_{si}$\\for all edges $s,i$};
	\node (sec2) [process, below of=Jp,node distance=1.9 cm] {Solve $\Delta T_{si}$\\in all edges $s,i$\\of sector 2};
	\node (sec1) [process, left of=sec2, node distance=4 cm] {Solve $\Delta T_{si}$\\in all edges $s,i$\\of sector 1};
	\node (dots) [empty, right of=sec2, node distance=2.5 cm] {...};
	\node (secn) [process, right of=dots, node distance=2.5 cm] {Solve $\Delta T_{si}$\\in all edges $s,i$\\of sector $n$};
	\node (parent) [process, below of=sec2] {Merge $\Delta T_{si}$ from overlapping sectors\\to parent object};
	\node (damp) [process, below of=parent,node distance=1.5 cm] {$\Delta T_{si}:=k\Delta T_{si}+(1-k)\Delta T_{{\rm p},si}$\\in all edges $s,i$};
	\node (update) [process, below of=damp,node distance=1.5 cm] {Update $\Delta\vJ$, $\Delta\vA$, $U$, $\vB$\\everywhere};
	\node (sectors) [process, below of=update,node distance=1.5 cm] {Copy all fields to sectors};
	\node (tol) [decision, below of=sectors, node distance=3.2 cm] {$\max|\Delta T_{si}-\Delta T_{{\rm p},si}|$\\$\le {\rm tol}$};
	\node (end) [startstop, below of=tol, node distance=3.5 cm] {End};
	\draw [arrow] (start) -- (Jinit);
	\draw [arrow] (Jinit) -- (Jp);
  \draw [arrow] (Jp.south) -- +(0.0,-0.25) -| (sec1);
  \draw [arrow] (Jp.south) -- +(0.0,-0.25) -| (sec2);
  \draw [arrow] (Jp.south) -- +(0.0,-0.25) -| (secn);
  \draw [arrow] (sec1.south) -- +(0.0,-0.4) -| (parent);
  \draw [arrow] (sec2.south) -- +(0.0,-0.4) -| (parent);
  \draw [arrow] (secn.south) -- +(0.0,-0.4) -| (parent);
	\draw [arrow] (parent) -- (damp);
	\draw [arrow] (damp) -- (update);
	\draw [arrow] (update) -- (sectors);
	\draw [arrow] (sectors) -- (tol);
	\draw [arrow] (tol) -- node[anchor=east] {yes} (end);
	\draw [arrow] (tol) -- node[anchor=south] {no} +(-6.0,0.0) |- (Jp);

\end{tikzpicture}
\caption{\E{The iterative parallel routine greatly reduces the computing time. The blocks in parallel solving $\Delta T_{si}$ at each sector represent the basic minimization routine of algorithm \ref{a.basic}. The tolerance ``tol" corresponds to $k|v|$, being $k$ the damping factor and $v$ the change in $\Delta T_{si}$ in algorithm \ref{a.basic}.}}
\label{f.parallel}
\end{figure}
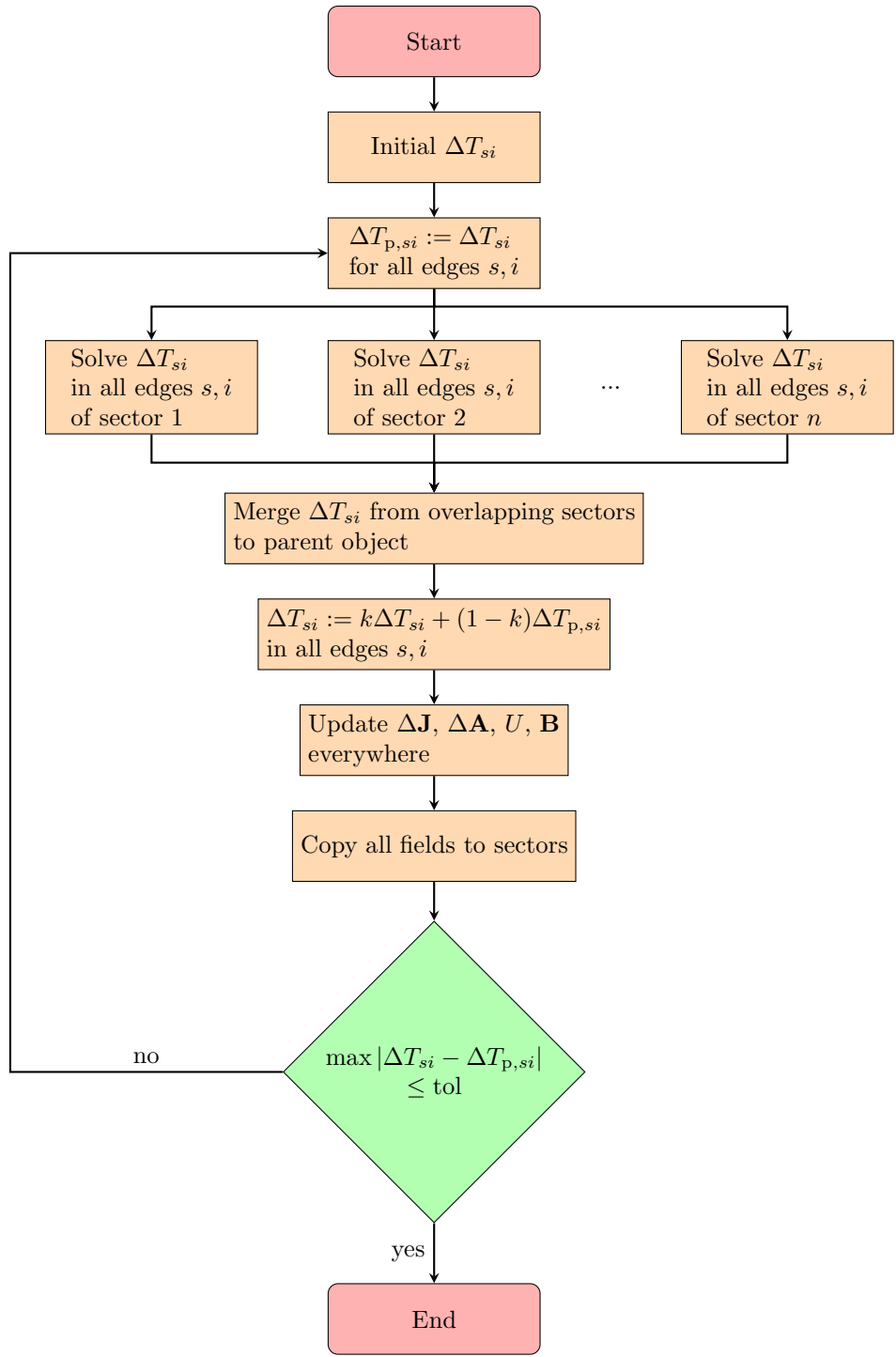

A general problem of 3D computations is the high number of degrees of freedom (DoF) required to achieve \E{sufficiently} accurate results. For instance, the calculations for the bulk sample in section \ref{s.constJc} use as many as 68921 cells, corresponding to 216972 DoF. This high number of DoF results in very long computing time. Here, we present a method to reduce the computing time without loss of accuracy.

\begin{figure}[ptb]
\centering
{\includegraphics[trim=-10 0 -10 0,clip,height=5.2 cm]{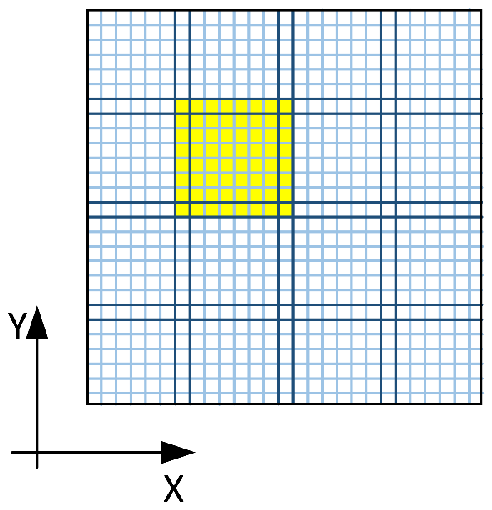}}
{\includegraphics[trim=-10 0 -10 0,clip,height=5.2 cm]{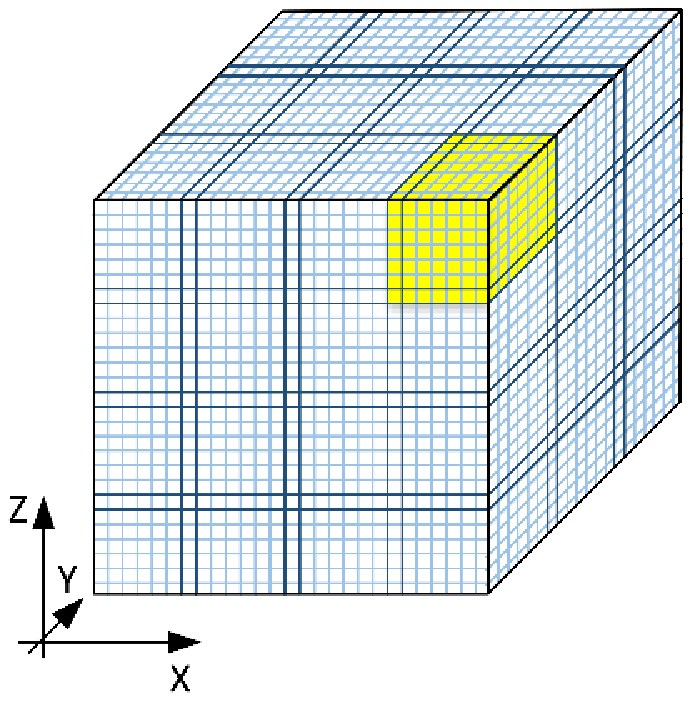}}
\caption{Dividing the sample into sectors speeds up the calculations with no loss of accuracy, in addition to enabling efficient parallelization. The left sketch is for a thin sample and the right is for a 3D object. For both cases, a particular sector is highlighted.}
\label{f.sectors}
\end{figure}

Similarly to \cite{magnet10k} for a cross-sectional method, we divide the sample into sectors. The main steps of the computation process are the following \E{(flux diagram in figure \ref{f.parallel})}.
\begin{enumerate}

\item The sample volume is divided into sectors, overlapping by a layer of one cell thickness (figure \ref{f.sectors}). Overlapping more than one cell may reduce the computing time, although this issue has not been explored in the present work.

\item $\Delta \vT$ is initialized to zero everywhere.

\item The program solves $\Delta \vT$ at each sector. \E{At the cell edges on the sector surface $\Delta\vT$ is not modified, keeping the value from the previous iteration}. After setting an increase in $\Delta \vT$ in one edge in the basic minimization routine, $\Delta\vA$ due to this increase is only updated within its own sector.

\item The solutions of each sector are merged into a single ``parent" object. Only $\Delta\vT$ at the edges within the sector volume are copied to the parent object, since the edges on the sector surface overlap with the neighboring sector. \E{In this way, all edges are modified in each iteration except those at the whole sample interface, where $\Delta\vT$ (and $\vT$) are kept as zero.} \E{A damping factor is applied in order to avoid oscillations.}

\item After merging, $\Delta\vJ$, $\Delta\vA$ and $U$ are updated in the whole parent object. This enables long-range magnetic interaction between sectors. 

\item \E{The values of $\Delta\vT$, $\Delta\vJ$, $\Delta\vA$ and $U$ are copied from the parent object to the sectors.}

\item The process from step \E{(3) to (6)} is repeated until the maximum difference in $\Delta\vT$ (or $\Delta\vJ$) between two iterations is below the same tolerance set by the basic minimization process within one sector. Thus, the division into sectors does not decrease the accuracy of the final result.

\end{enumerate}

The advantages of the division into sectors are two-fold. First, the routine can be efficiently parallelized, essential for multi-core processors and computer clusters. Second, the magnetic field created by one sector decreases at least as power 2 with the distance, with increasing power for multipole contributions of increasing order. This causes that coarse solutions in one sector generate sufficiently accurate magnetic fields in another distant sector, limiting both the number of iterations and the computing time of each iteration. Since the computing time scales as power 2 of the DoF, the computing time of one iteration for $m$ sectors is $1/m$ of that of the whole object. For example, dividing the volume into 1000 sectors, a problem requiring around 10 iterations will reduce the computing time by roughly factor 100.

The computing time is optimized if the computation is done first for a coarse tolerance and we repeat the whole process by decreasing it by factor 10, repeating the process until we reach the goal tolerance.

\E{For magnetic-field dependent parameters in the $\vE(\vJ)$ relation, we do not need to apply the iterations mentioned in section \ref{s.min} when solving each sector separately. The reason is that the whole process is already iterative, being $\vB$ evaluated at each iteration (figure \ref{f.parallel}).}

In this work, we implemented the parallel minimization routine in C++ using the OpenMP protocol.

\subsection{Symmetries and computing time}
\label{s.sym}

With the division into sectors, symmetries can be taken into account straightforwardly. The effective magnetization $\vT$ is computed in one eighth of the rectangular prism or one fourth of the 2D rectangle. When importing to the parent object, $\vT$ is copied from the computed region to the rest of the sample. The update of $\vA$ is made on the whole body. In this way, the computing time can be reduced by $1/8$ and $1/4$ for rectangular prisms and films, respectively. 

The computing time of the cube (216972 DoF) in figure \ref{cubeJc.fig} is of 14 hours in a computer with a 4-core (8 threads) processor Intel Core i7-4771 and 8 Gb RAM.

\subsection{Magnetization and AC loss}
\label{s.mag}

The average magnetization ${\bf M}$ is defined as the total \E{magnetic moment} $\bf m$ per unit sample volume $V$ as ${\bf M}={\bf m}/V$. The magnetic moment is
\begin{eqnarray}
{\bf m  } & = & \half\int\dif V\ \vr\times\vJ \label{mrJ} \\
& = & \int\dif V\ \vT.
\end{eqnarray}
We evaluate the integral in (\ref{mrJ}) by assuming $\vr\times\vJ$ uniform in the cells and taking the values at the cells center, being $\vJ$ interpolated there.

The local instantaneous AC loss is $\vE\cdot\vJ$ \cite{acreview}, and hence the instantaneous power loss is
\begin{equation}
P=\int \dvol\ \vE(\vJ)\cdot\vJ.
\end{equation}
For our discretization, $\vE(\vJ)\cdot\vJ$ is assumed uniform within the cells and, again, we take $\vJ$ as the interpolated value.


\section{Model tests}
\label{s.tests}

This section tests the model with analytical limits. First, we compare our model with Halse's analytical formula for an infinite strip \cite{halse70JPD} under uniform applied magnetic field. Then the magnetization of a thin disk is checked again with the analytical formula in \cite{clem94PRB}. In both cases, the computations agree very well with the analytical limits, supporting the correctness of the model.

\subsection{Thin strip}

\begin{figure}[ptb]
\centering
\subfloat[][]
{\includegraphics[trim=-10 0 -10 0,clip,width=4.5 cm]{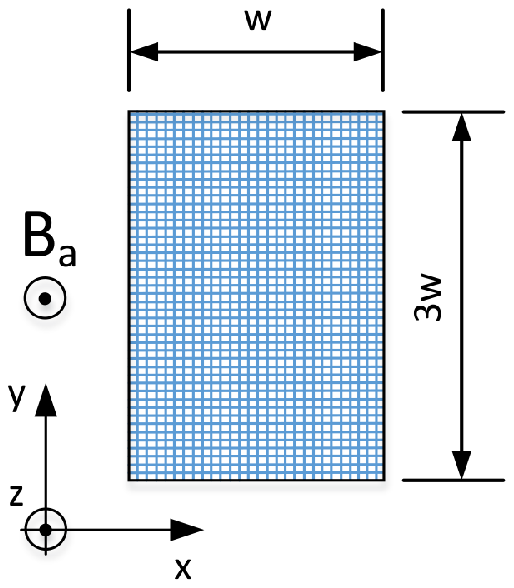}}
\subfloat[][]
{\includegraphics[trim=-10 0 -10 0,clip,width=4.5 cm]{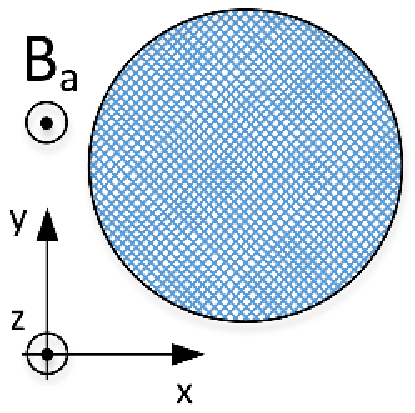}}
\subfloat[][]
{\includegraphics[trim=0 -20 0 0,clip,width=4.5 cm]{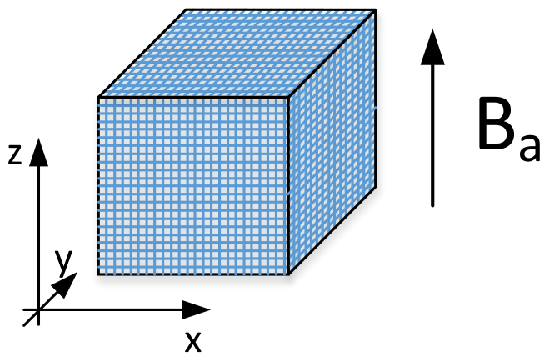}}
\caption{Sketch of the thin film (a), thin disk (b), and cube (c) under uniform applied magnetic field. Dimensions are in mm, being $R$ is the disk radius.}
\label{film.fig}
\end{figure}

Here, we consider a long thin strip, such as that in figure \ref{film.fig}a. Our computations assume constant ${J_{c}}$ dependence and isotropic power law with ${n}$-factor 1000. The analytical formula is based on the CSM, and hence we use a very high $n$-factor to approach the smooth $\vE(\vJ)$ relation as much as possible to that of the CSM. The sinusoidal applied field ${B_{a}}$ is parallel to the ${z}$ axis with amplitude 20 mT and frequency 50 Hz. The dimensions of the computed sample, width $\times$ length $\times$ thickness, are ${4\times 12 \times 10^{-3}}$ mm$^3$. The total number of cells is 34347, distributed as ${107\times321\times1}$. The critical current density is ${J_{c}=2.72\cdot10^{10}}$ A/m${^2}$, which is similar to that of common commercial tapes.

\begin{figure}[ptb]
\centering
{\includegraphics[trim=0 0 0 0,clip,width=5.5 cm]{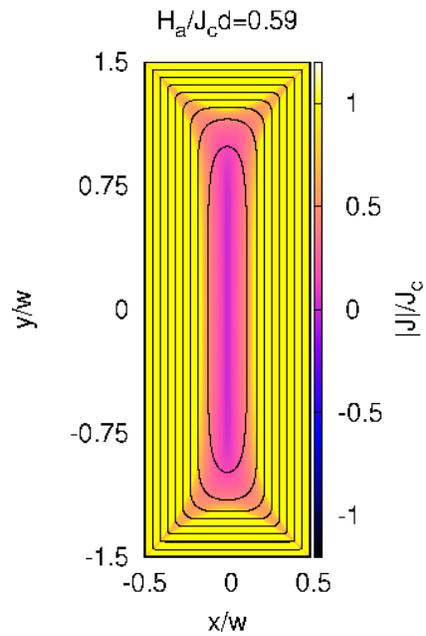}}
\caption{Current flux lines and modulus of the current density (colormap) at the peak of the AC applied magnetic field. The power-law exponent is taken as 1000. The current density follows the ${y}$ direction far away from the ends.}
\label{Brant1.fig}
\end{figure}

\begin{figure}[ptb]
\centering
{\includegraphics[trim=0 0 0 0,clip,width=8.5 cm]{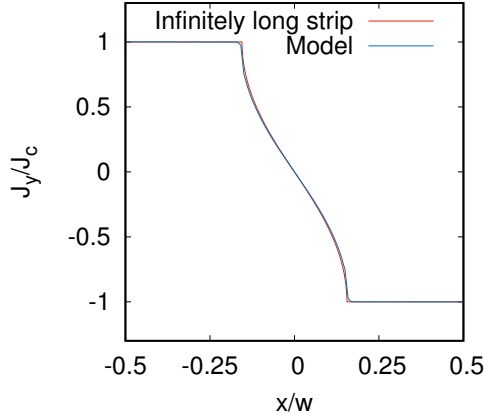}}
\caption{The computed $J_y$ at the midplane ($y=0$) for power-law exponent 1000 agrees with the thin strip formula \cite{halse70JPD}.}
\label{Brant2.fig}
\end{figure}

The formula for the current density in a long strip at the initial magnetization stage, from zero applied field to the peak, is \cite{halse70JPD,Brandt93PRBa,Zeldov94PRB}
\begin{eqnarray}
 J_{y}(x) & = & \frac{2J_{c}}{\pi}\arctan{ \frac{cx}{\sqrt{(b^{2}-x^{2})}}},\qquad  |x|<b,  \nonumber \\ 
          & = & J_{c}\frac{x}{|x|},\qquad   b<|x|<w/2,   \label{Brant1} \nonumber \\
\end{eqnarray} 
where 
\begin{equation}
{b=\frac{w}{2\cosh{\frac{H_{a}}{H_{c}}}}},   \label{} \\
\end{equation}  
\begin{equation}
{c=\tanh{\frac{H_{a}}{H_{c}}}},   \label{} \\
\end{equation}  
and
\begin{equation}
{H_{c}=\frac{J_{c}d}{\pi}}.   \label{}
\end{equation}
The other parameters are the thickness ${d}$, the width $w$, and the applied magnetic field ${H_a=B_a/\mu_{0}}$. The screening current at the peak of applied field is shown on figure \ref{Brant1.fig}. The current density profile at the peak of the applied field for the long strip formula (\ref{Brant1}) is compared to the numerically computed one at the central plane, defined as $y=6$ mm. As seen in figure \ref{Brant2.fig}, our model agrees with the analytical formula very well.

\subsection{Thin disk}

We also test MEMEP for the magnetization of a thin disk, such as that of figure \ref{film.fig}b. The disk radius in the calculations is ${R=6}$ mm and the thickness is $d=1$ ${\mu}$m. The applied field follows the $z$ axis and is of 8.00 mT amplitude and 50 Hz frequency. The critical current density is ${2.72\cdot10^{10}}$ A/m${^{2}}$ and we again take an $n$-factor of 1000. We used the analytical formula in \cite{clem94PRB} for the thin disk. The current density at the initial magnetization stage is
\begin{eqnarray}
J_{y}(x) & = & \frac{-2J_{c}}{\pi}\arctan{ \frac{\frac{x}{R}\sqrt{(R^{2}-a^{2})}}{\sqrt{(a^{2}-x^{2})}}},\qquad  x\leq a,  \nonumber \\ 
         & = & -J_{c},\qquad   a\leq x<R,   \label{Disk}
\end{eqnarray} 
where 
\begin{equation}
{a=\frac{R}{\cosh{\frac{H_{a}}{H_{d}}}}}   \label{} 
\end{equation}  
and
\begin{equation}
{H_{d}=\frac{J_{c}d}{2}}.   \label{}
\end{equation}

\begin{figure}[ptb]
\centering
{\includegraphics[trim=0 0 0 0,clip,width=8.5 cm]{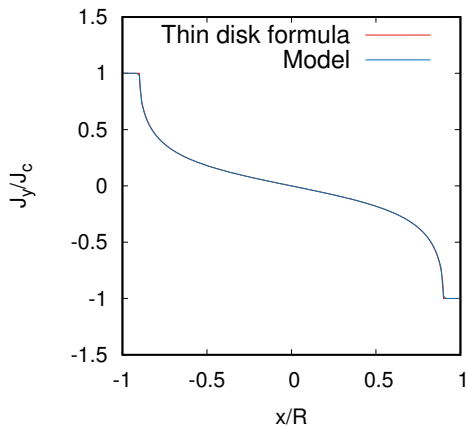}}
\caption{For a thin disk, the calculated $J_y$ at the midplane ($y=0$) for a power-law exponent 1000 agrees well with the analytical predictions \cite{clem94PRB}.}
\label{disk0.fig}
\end{figure}

\begin{figure}[ptb]
\centering
{\includegraphics[trim=0 0 0 0,clip,width=8.5 cm]{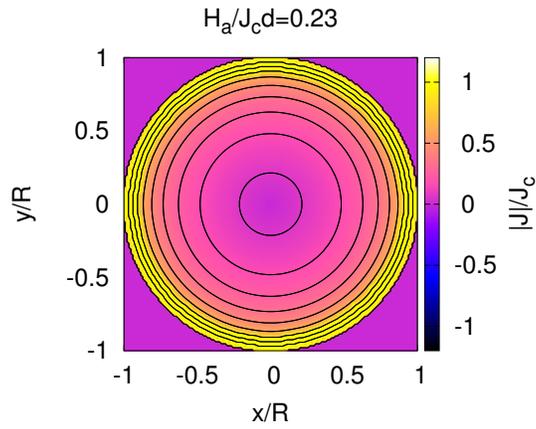}}
\caption{For a thin disk, the computed current flux lines and current density magnitude (colormap) show that the magnetization currents flow in circular loops, while cylindrical symmetry is not imposed.}
\label{disk1.fig}
\end{figure}

In figure \ref{disk0.fig}, we compare the computed current profile at 8.00 mT applied field at the initial magnetization curve. The current density calculated by MEMEP agrees very well with the formula. As seen in figure \ref{disk1.fig}, the solution of the computed current density follows cylindrical symmetry within numerical error, although the model does not impose such symmetry. Indeed, the applied vector potential of (\ref{Aalong}) lacks cylindrical symmetry, following the $y$ direction.

We also compare the results for the hysteresis loop. The analytical formula \cite{clem94PRB} of magnetization hysteresis in disks is split into 3 functions, corresponding to the initial curve, and decreasing and increasing applied magnetic fields, respectively. The initial curve is
\begin{equation}
{M_{zi}(H_{a})=-\chi_{0}H_{a}S(H_{a}/H_{d})},   \label{} 
\end{equation} 
with
\begin{eqnarray}
&& {\chi_{0}=\frac{8R}{3\pi d}}   \label{} \\
&& {S(x)=\frac{1}{2x}\left[ \cos^{-1} \left (\frac{1}{\cosh x} \right ) + \frac{\sinh x}{\cosh^{2} x}\right]}.   \label{}
\end{eqnarray}
The two remaining functions for decreasing and increasing applied magnetic fields, respectively, are 
\begin{equation}
{M_{z\downarrow}=M_{zi}(H_{m})-2M_{zi}(\frac{H_{m}-H_{a}}{2})},   \label{}
\end{equation} 
\begin{equation}
{M_{z\uparrow}=-M_{zi}(H_{m})+2M_{zi}(\frac{H_{a}-H_{m}}{2})},   \label{}
\end{equation} 
where $H_m$ is the amplitude of the applied magnetic field. 

\begin{figure}[ptb]
\centering
{\includegraphics[trim=40 0 25 0,clip,width=8.5 cm]{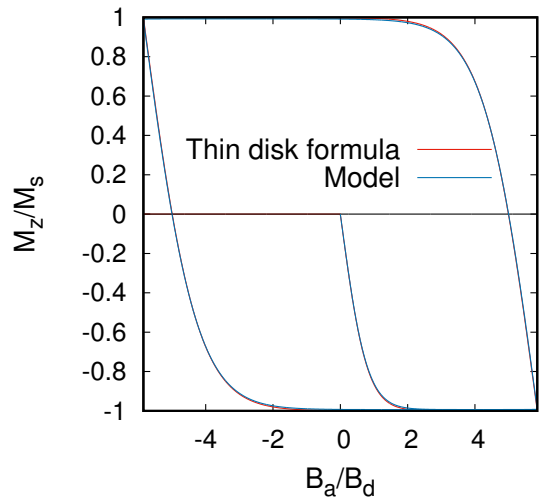}}
\caption{The computed magnetization loop for the thin disk agrees with the analytical formulas in \cite{clem94PRB}.}
\label{diskmag.fig}
\end{figure}

The comparison of the previous analytical formulas and the model is at figure \ref{diskmag.fig}. The model agrees very well with the analytical limit. The model uses 80 time steps per cycle.

\section{Examples and discussion}
\label{s.examp}

This section presents the current density for a cube of side $w$. We consider both constant $J_c$ and Kim-like $J_c({\bf B})$ dependence. For all cases, we consider a power-law exponent of 100. The high power-law exponent ensures that the solution is representative of the CSM.


\subsection{Current density with constant $J_c$}
\label{s.constJc}

Here we analyze in detail a cubic bulk sample with applied field perpendicular to one side, as in the sketch in figure \ref{cubeJc.fig}. In particular, we consider a sample of ${10\times10\times10}$ mm$^3$, $J_c=10^8$ A/m$^2$ and power-law exponents of $n=$100. The applied magnetic field follows the $z$ direction and is of 200 mT amplitude and 50 Hz frequency. 

The magnetization current density at the peak of applied field for $n=100$ is at figures \ref{cubeJc.fig} and \ref{cubecJy.fig}. The main screening currents are flowing in closed loops perpendicular to the applied field (figures \ref{cubeJc.fig}abc). The highest penetration depth is at the top and bottom (figure \ref{cubeJc.fig}c), close to the surface, being the smallest at the middle (figure \ref{cubeJc.fig}a). The ${J_{y}}$ component of current density shows the penetration depth in the entire cross-section of the cube at the mid-plane defined by ${y}$=5 mm (figure \ref{cubecJy.fig}a).

Up to now, the current penetration is qualitatively similar to cylinders \cite{brandt98PRBa,sanchez01PRB}. However, the cube presents non-zero $J_z$ component (figures \ref{cubeJc.fig}de), which reaches values as high as 30 \% of ${J_{c}}$. The highest magnitude of ${J_{z}}$ is close to the diagonal of the cube (figure \ref{cubeJc.fig}e). This $J_z$ bends the current flux lines, as seen in the section close to the lateral surface of figure \ref{cubeJc.fig}d and the 3D current loop in figure \ref{Jline3d.fig}. The cause of this $J_z$ component is the self-field. In cylinders, the radial component of the self-field, perpendicular to the current loops, is balanced by higher current penetration close to the ends \cite{sanchez01PRB}. That is possible thanks to the cylindrical symmetry, which causes that the radial field is uniform in any circular loop. This no longer applies to rectangular prisms. The magnetic field created by rectangular loops at the diagonal is higher than closer to the straight parts at the same distance from the lateral faces \cite{brandt95PRL}. Thus, higher current penetration close to the ends following rectangular loops cannot fully cancel the self-field. Close to the diagonals, the additional perpendicular self-field pointing inwards is canceled by a $J_z$ component that changes its sign at the diagonal. For applied fields well above the penetration field, the self-field is not relevant, and hence the current paths follow rectangular loops in the whole sample (figure \ref{cubeJc2Bp.fig}).

The presence of the non-zero $J_z$ component contrasts with earlier predictions in \cite{badia05APL}, where in-plane square loops were assumed for a cube. This assumption was supported by taking into account that $|J|$ follows $|J|=J_c$ or 0 only, while the CSM allows any $|J|\le J_c$. Thus, the discrete symmetries of the cube are not sufficient to impose square current loops. Current densities with magnitude slightly below $J_c$ are enough to bend $\vJ$ vertically and obtain the necessary $J_z$ to shield the self-field. Current densities with regions of $|J|<J_c$ have also been shown in \cite{brandt95PRBa,prigozhin98JCP} for thin films, presenting non-square current paths. Nevertheless, the assumption of square loops should still provide a good approximation of the magnetic moment.

\begin{figure}[ptb]
\centering
{\includegraphics[trim=0 0 0 0,clip,width=12 cm]{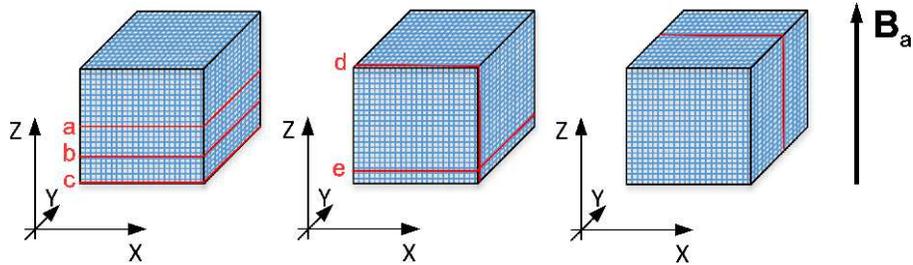}}
\caption{\E{The left and center sketches show the planes of the (a,b,c) and (d,e) plots, respectively, in figures \ref{cubeJc.fig}, \ref{cubeJc2Bp.fig} and \ref{cubeJc(B).fig}. The right sketch shows the plane of the plots in figures \ref{cubecJy.fig} and \ref{cubeJc(B)Jz.fig}. }}
\label{planes.fig}
\end{figure}

\begin{figure}[ptb]
\centering
{\includegraphics[trim=0 -10 0 0,clip,height=4.2 cm]{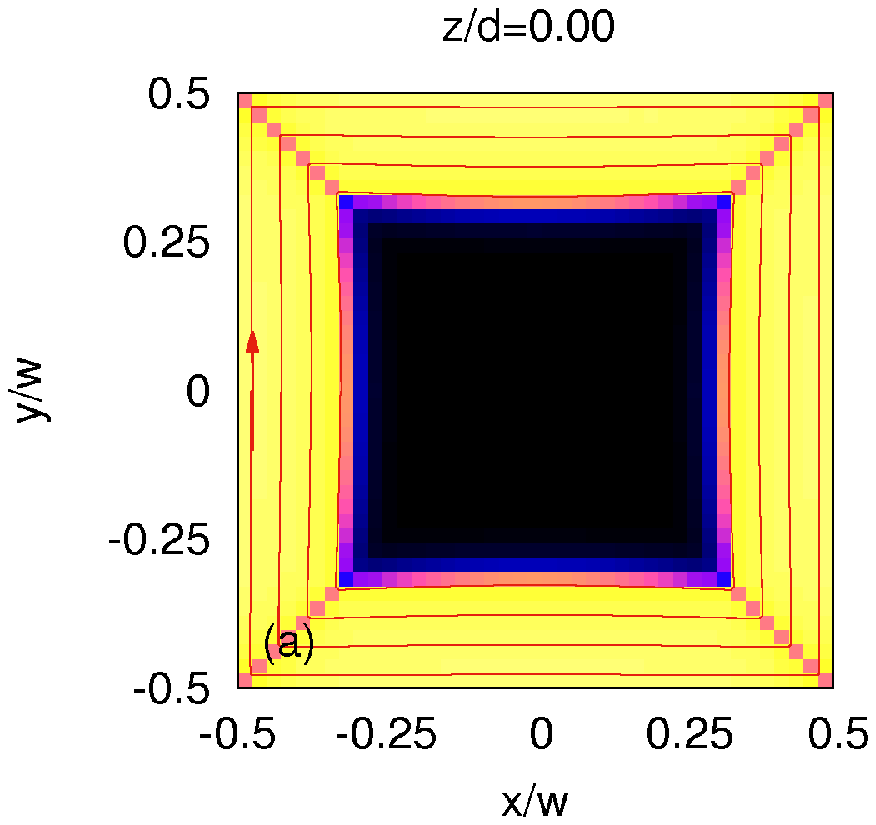}}
{\includegraphics[trim=0 -10 0 0,clip,height=4.2 cm]{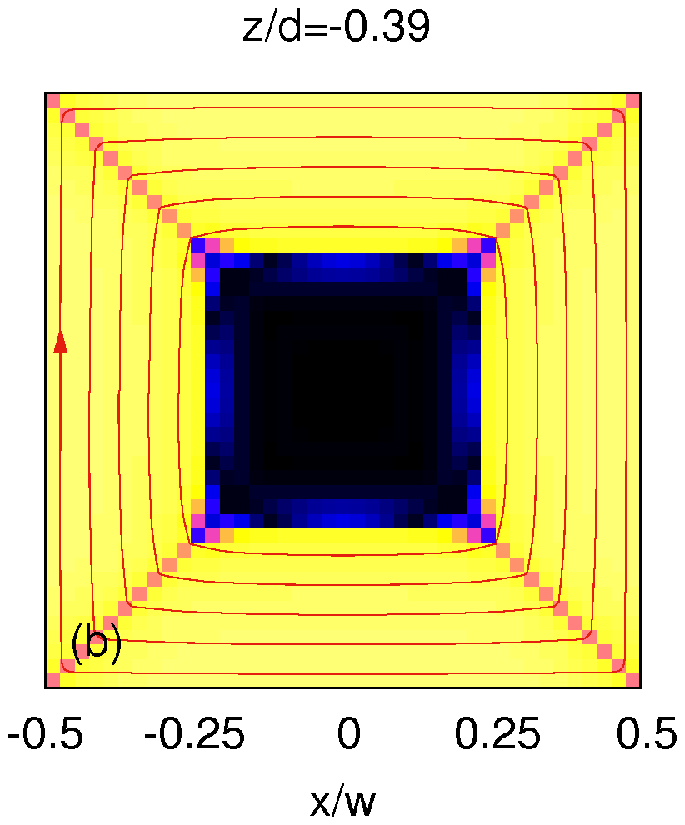}}  
{\includegraphics[trim=0 -10 0 0,clip,height=4.2 cm]{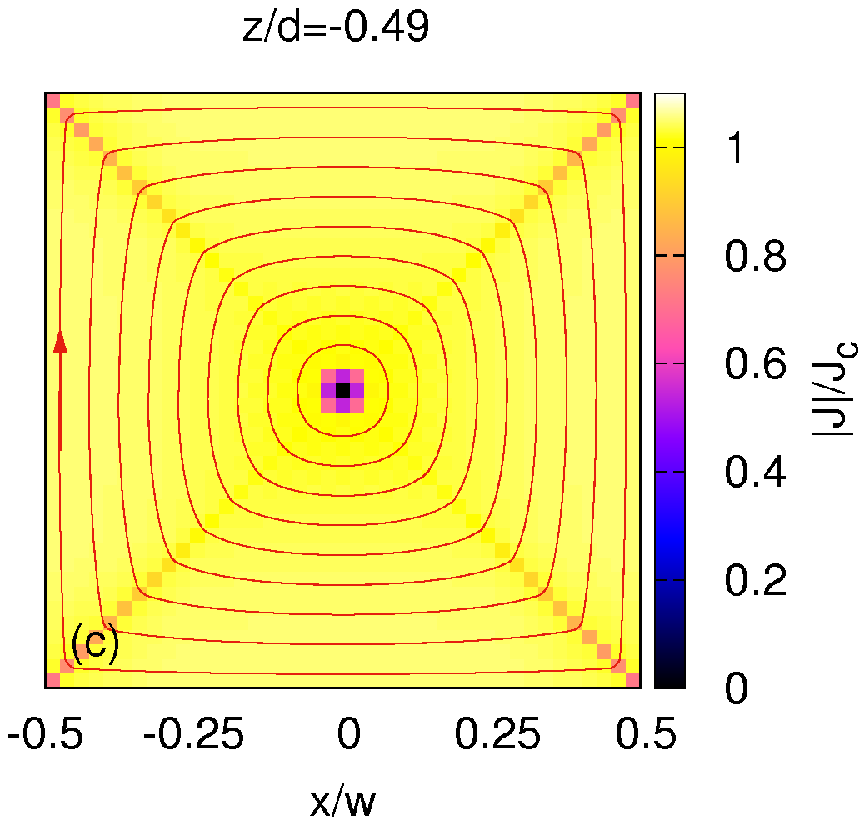}} \\
{\includegraphics[trim=0 0 0 0,clip,height=4.2 cm]{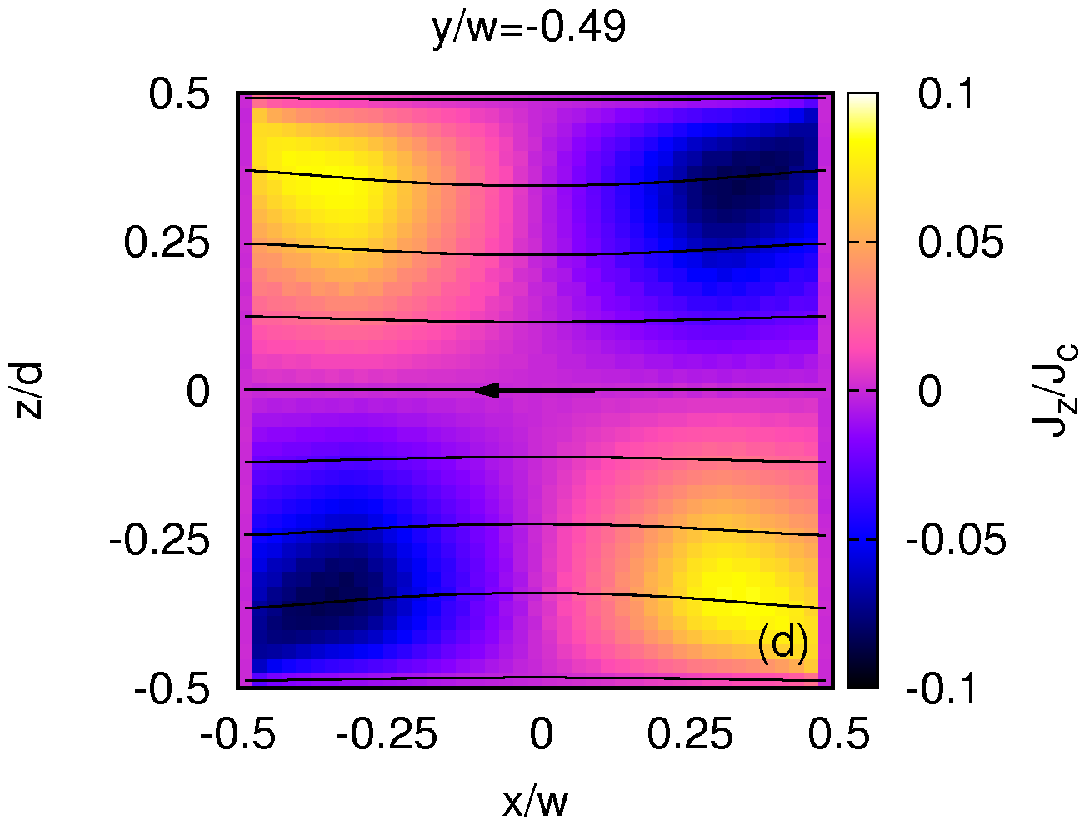}}
{\includegraphics[trim=0 0 0 0,clip,height=4.2 cm]{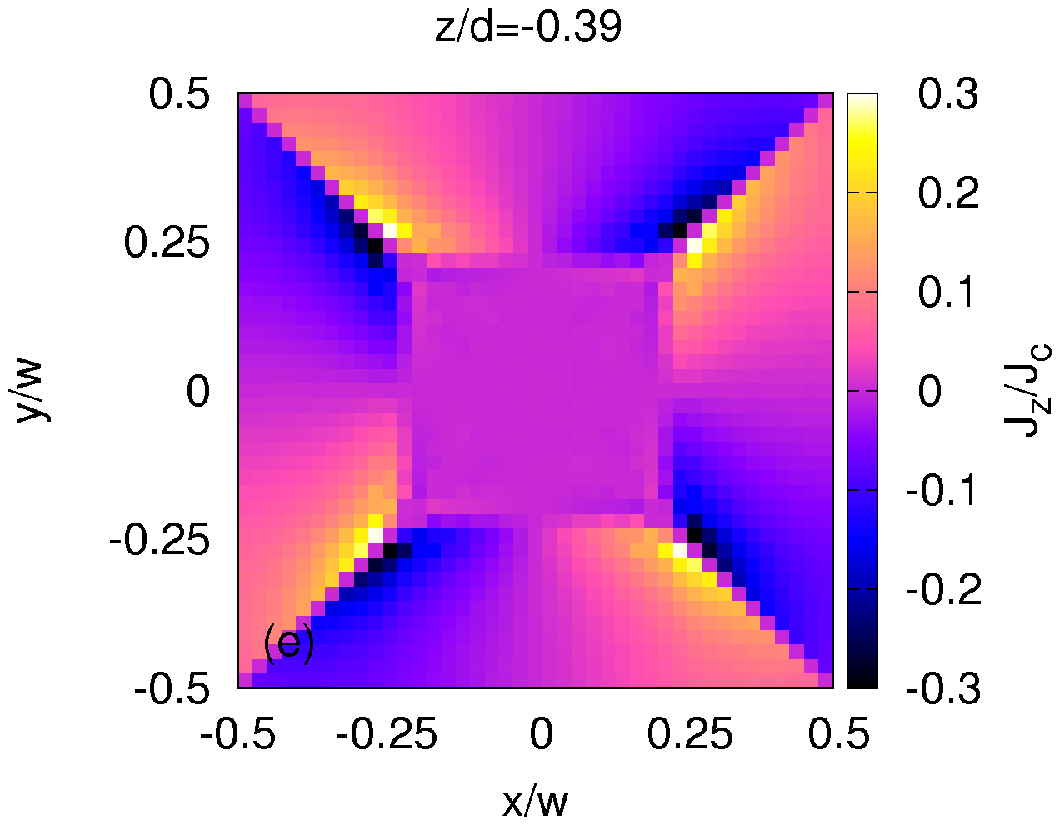}}
\caption{Current density magnitude (a,b,c) and $J_z$ (d,e) at several cross-sections of a cube ($d/w=1$), \E{corresponding to planes (a,b,c,d,e) in figure \ref{planes.fig}}. The lines are 3D current flux lines projected on the plotted plane that start at $y=0$ in (a,b,c) and $x=0.5w$ in (d), representing the direction of the current density but not its magnitude. $J_z$ in (d) is for the $z$ plane where $|J_z|$ is the highest, $z/d=0.11$. Computed case for constant $J_c$ and applied field $B_a=0.155 J_cw$.}\label{cubeJc.fig}
\end{figure}

\begin{figure}[ptb]
\centering
{\includegraphics[trim=0 0 0 0,clip,height=5.0 cm]{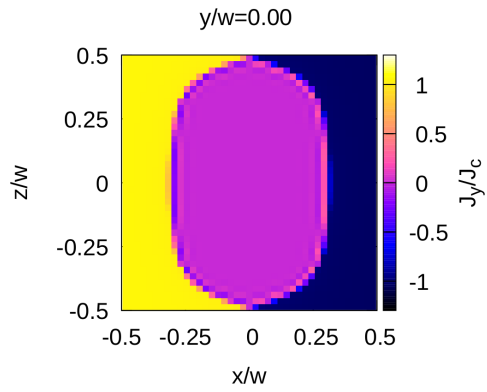}}
\caption{$J_y$ at the $y=0$ midplane \E{(right sketch in figure \ref{planes.fig})} of a cube under the same situation as figure \ref{cubeJc.fig}.}\label{cubecJy.fig}
\end{figure}

\begin{figure}[ptb]
\centering
{\includegraphics[trim=10 20 25 20,clip,width=10.0 cm]{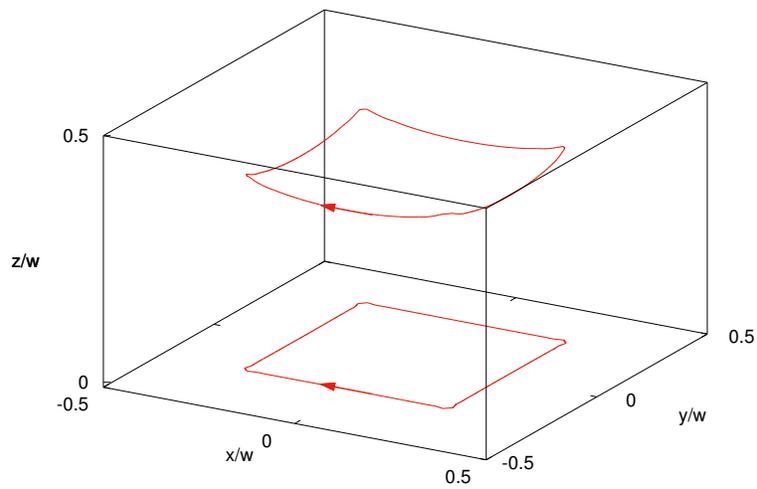}}
\caption{This example of 3D current flux lines in the cube of figure \ref{cubeJc.fig} shows that the current lines in part of the cube present off-plane bending (upper flux line), while close to the center the flux lines are square (lower flux line).}
\label{Jline3d.fig}
\end{figure}

\begin{figure}[tbp]
\centering
{\includegraphics[trim=0 -10 0 0,clip,height=4.2 cm]{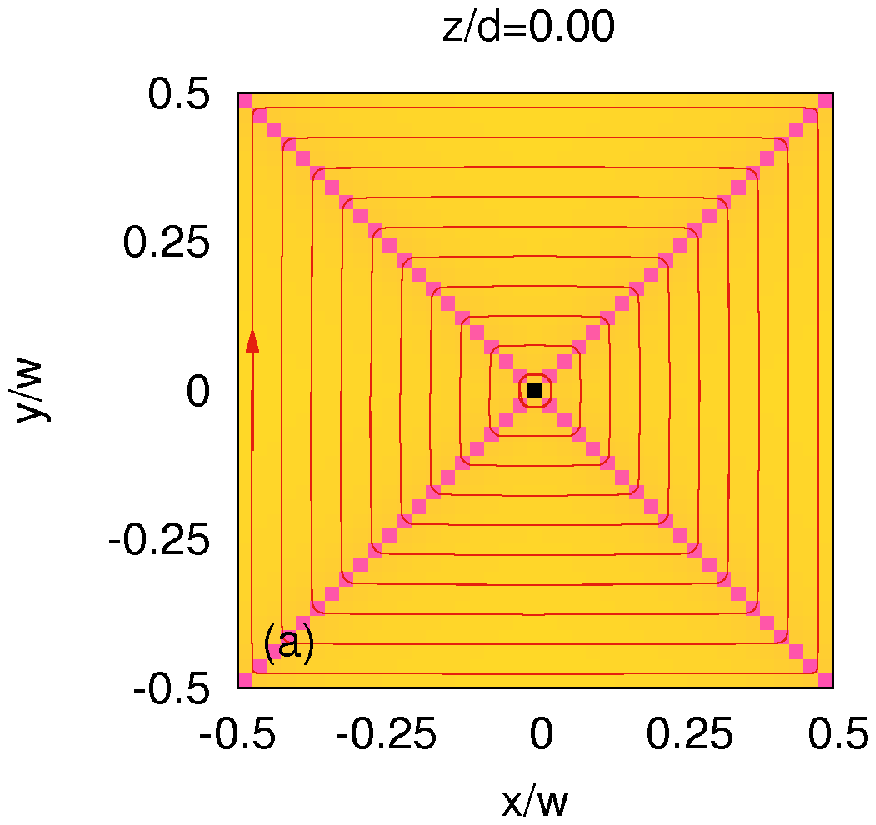}}
{\includegraphics[trim=0 -10 0 0,clip,height=4.2 cm]{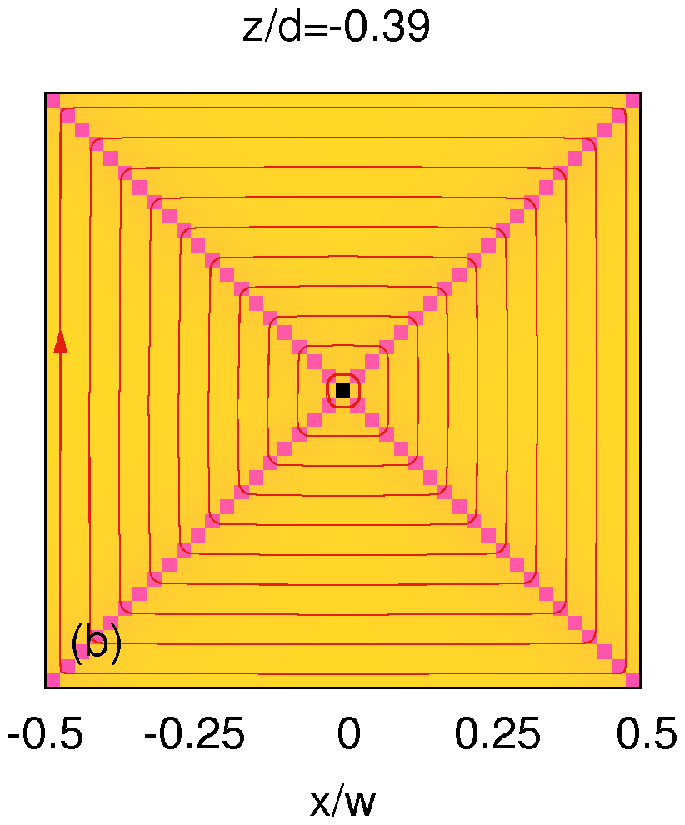}}  
{\includegraphics[trim=0 -10 0 0,clip,height=4.2 cm]{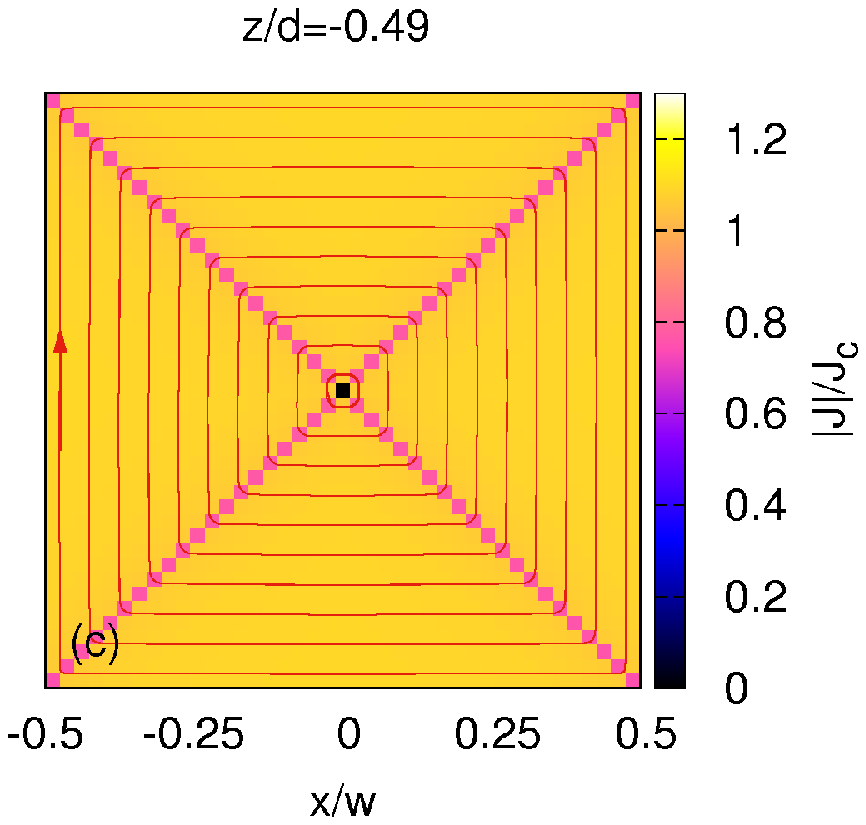}} \\
{\includegraphics[trim=0 0 0 0,clip,height=4.2 cm]{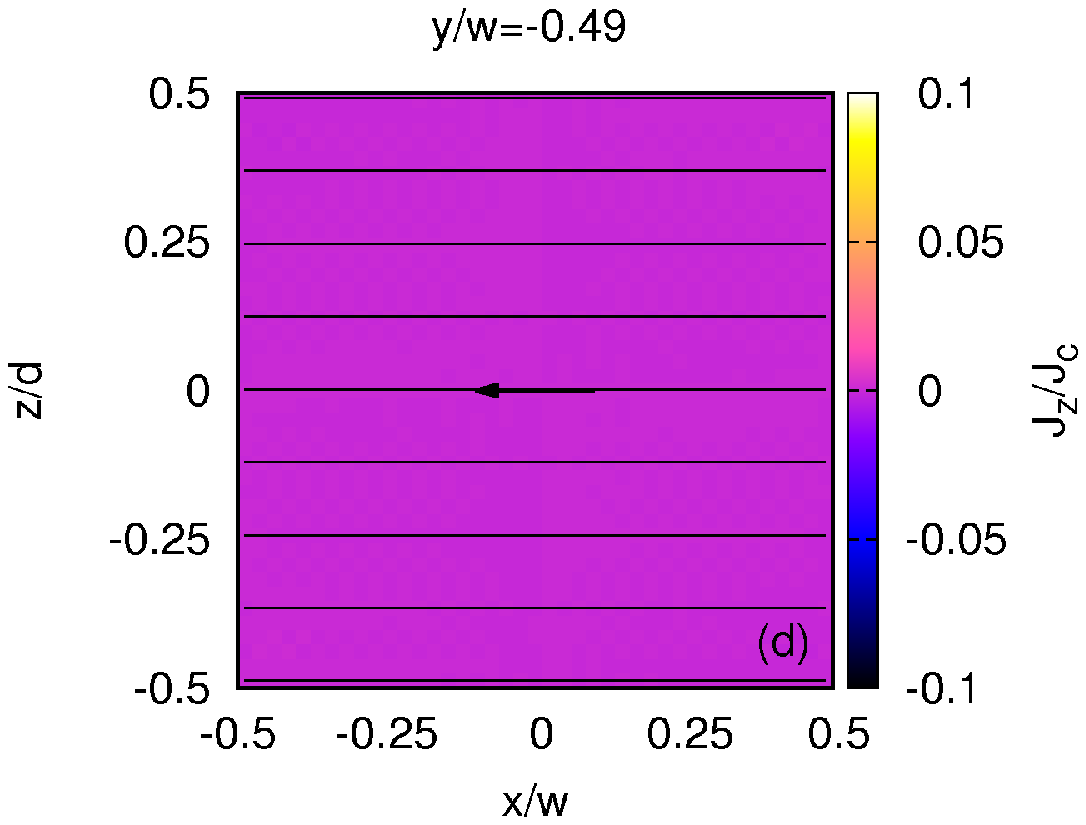}} 
{\includegraphics[trim=0 0 0 0,clip,height=4.2 cm]{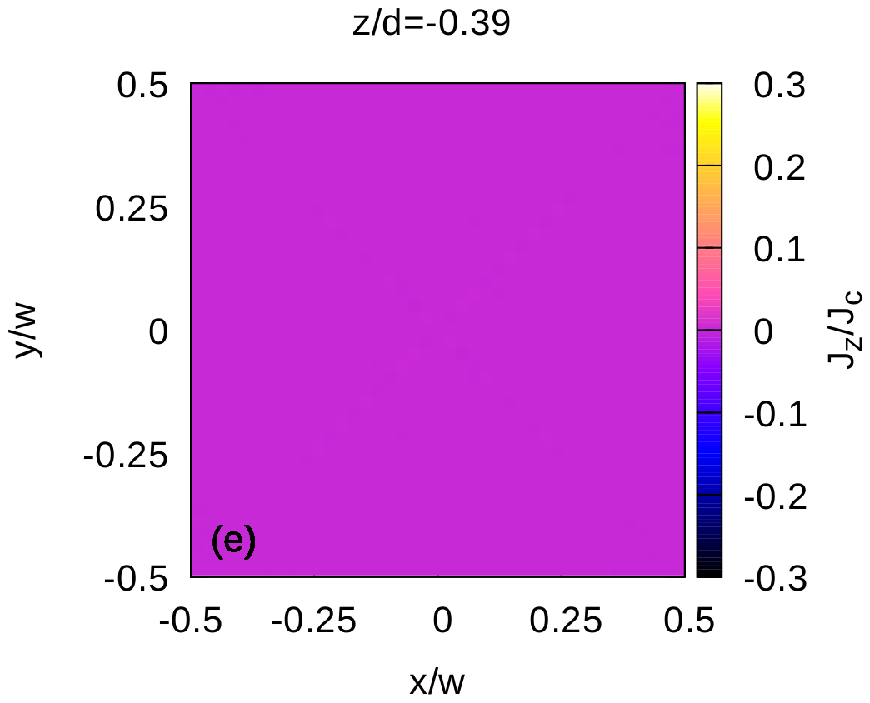}}
\caption{The same as figure \ref{cubeJc.fig} but for an applied field $B_a=0.310J_cw$ \E{[maps for the planes (a,b,c,d,e) in figure \ref{planes.fig}]}. At high applied fields, the current lines are square and $J_z$ vanishes, as in CSM predictions for long bars.}\label{cubeJc2Bp.fig}
\end{figure}


\subsection{Current density with magnetic-field dependent $J_c$} 
\label{s.JcB}

Here, we assume a ${J_c(\vB)}$ dependence according to Kim's \E{formula}, with constants ${J_{c0}=10^{8}}$ A/m${^{2}}$ and ${B_{0}=}$20 mT. We consider a cube with the same geometry and AC applied magnetic field as the previous one for constant $J_c$. We take a power-law exponent of $100$.

\begin{figure}[ptb]
\centering
{\includegraphics[trim=0 -10 0 0,clip,height=4.2 cm]{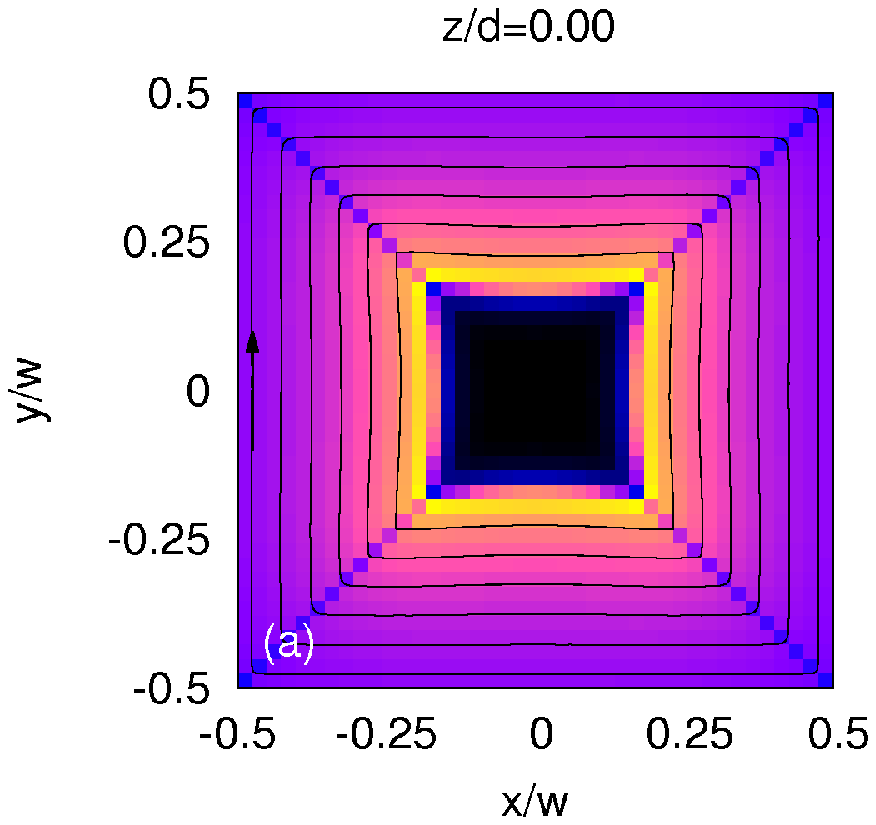}}
{\includegraphics[trim=0 -10 0 0,clip,height=4.2 cm]{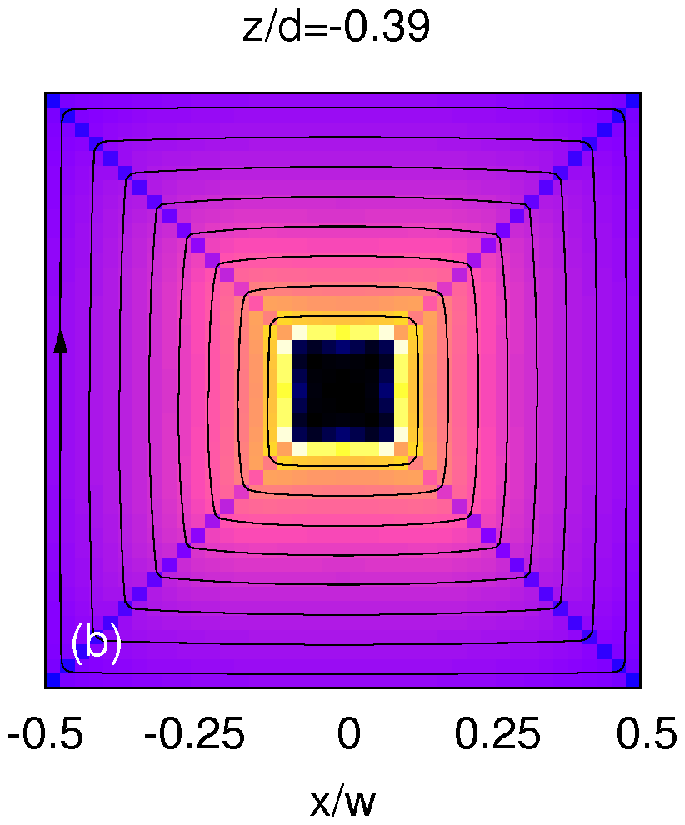}}  
{\includegraphics[trim=0 -10 0 0,clip,height=4.2 cm]{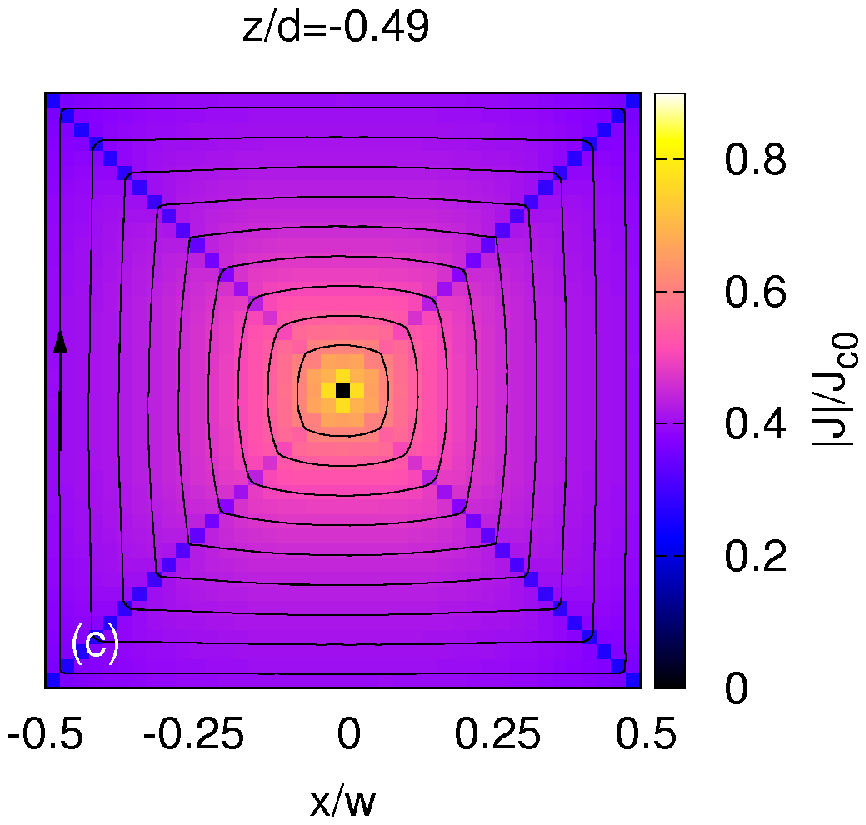}} \\
{\includegraphics[trim=0 0 0 0,clip,height=4.2 cm]{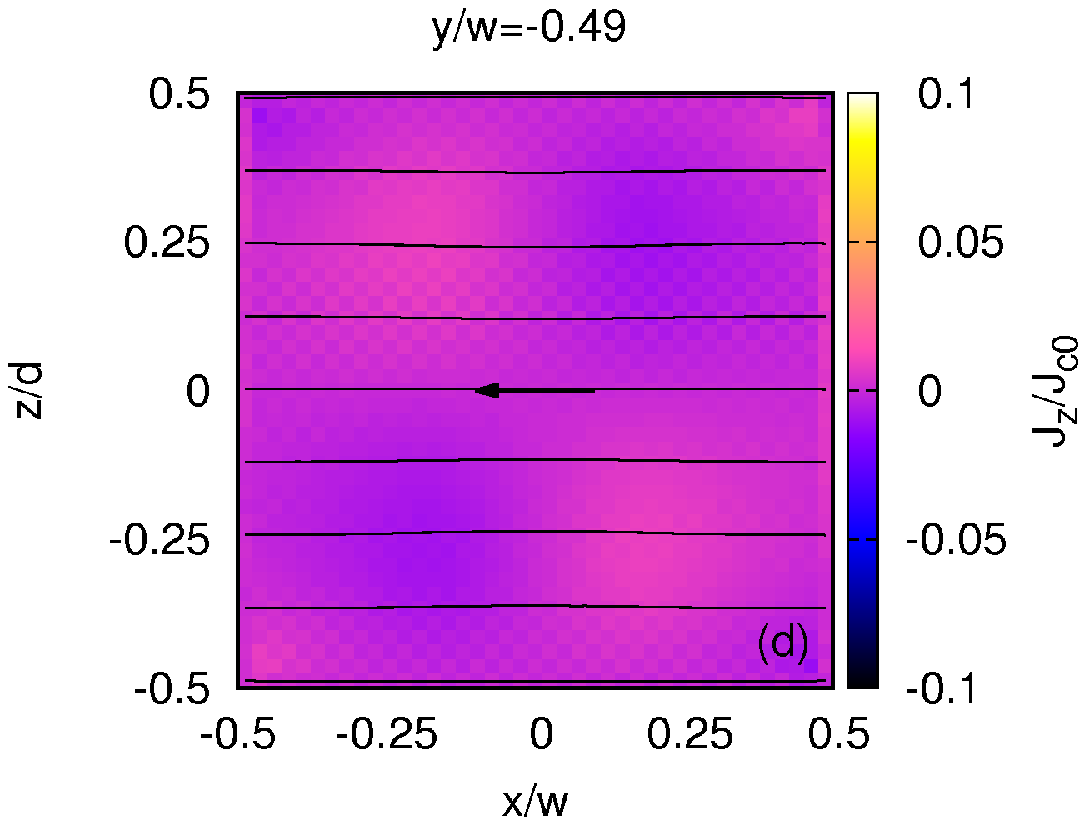}} 	
{\includegraphics[trim=0 0 0 0,clip,height=4.2 cm]{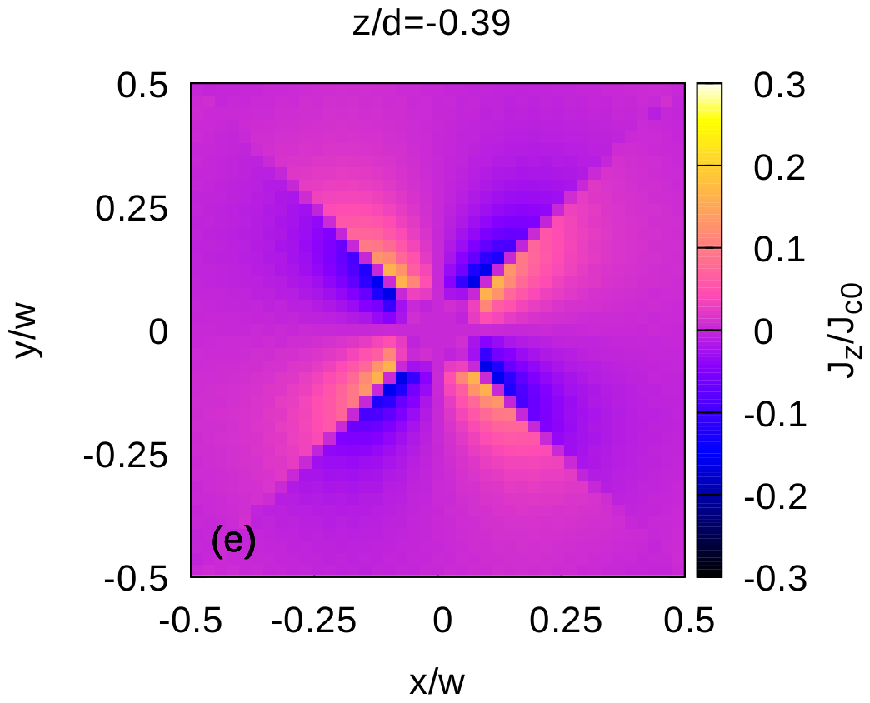}}
\caption{The same as figure \ref{cubeJc.fig} but for a cube with Kim-like magnetic-field-dependent $J_c$ (rest of parameters in the text). \E{Maps for the planes (a,b,c,d,e) in figure \ref{planes.fig}.}}\label{cubeJc(B).fig}
\end{figure}

\begin{figure}[ptb]
\centering
{\includegraphics[trim=0 0 0 0,clip,height=5 cm]{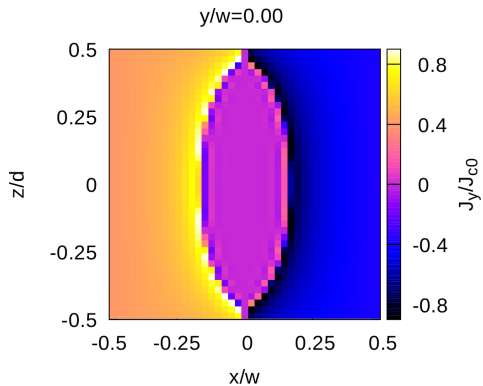}} 
\caption{The same as figure \ref{cubecJy.fig} but for for a cube with Kim-like magnetic-field-dependent $J_c$ (rest of parameters in the text). \E{Plane of the map in figure \ref{planes.fig}.}}\label{cubeJc(B)Jz.fig}
\end{figure}

Figure \ref{cubeJc(B).fig} shows the current penetration for the instantaneous applied magnetic field of ${B_{a}=}$178 mT. The screening current mainly presents the same behaviour as for constant $J_c$, with the difference that $|\vJ|$ is higher at the border with the current-free region than at the cube surface; where $|\vJ|\approx$ $J_{c0}$ and 0.5$J_{c0}$, respectively (see figures \ref{cubeJc(B).fig}a and \ref{cubeJc(B)Jz.fig}). This is caused by the magnetic-field dependence of $J_c$, since $|\vB|$ vanishes at the current-free core and is the largest at the cube surface. The sample with $J_c(\vB)$ is closer to saturation than that with constant $J_c$. This causes that $J_z$ vanishes in an important portion of the sample and reduces the maximum $J_z$, being only 0.2$J_{c0}$ (figure \ref{cubeJc(B).fig}d). The current flux lines are almost square close to the surface of the sample (figure \ref{cubeJc(B).fig}c). This is caused by the relatively low importance of the self-field in most of the cross-section.

\section{Conclusions}
\label{s.concl}

This article presented a novel formulation of a variational principle for 3D modeling of non-linear eddy currents, such as those present in superconductors. As example, we analyzed a cube under uniform applied magnetic field.

The variational principle, which enables to reduce the computation volume to the sample, takes the effective magnetization $\vT$ as state variable in order that the magnetization current density is $\nabla\times\vT$. This formulation is also valid if the sample is submitted to a transport current, in addition to the applied magnetic field. We have shown that the physical solution minimizes the functional and that the minimum is unique. 

Our implementation uses an original time-efficient minimization method. The computing time is enhanced by dividing the sample volume into sectors and solving all sectors iteratively with no loss of accuracy. This routine has also been efficiently parallelized. Tests with thin films and cylinders showed perfect agreement with existing analytical formulas.

We computed and analyzed $\vJ$ of a superconducting cube under uniform applied magnetic field. We found both non-zero component of the current density parallel to the applied field and non-square current paths close to the top and bottom of the sample. The cause of both phenomena is the self-field, disappearing for applied fields well above the penetration field. Although the results are for a power-law $\vE(\vJ)$ relation, these phenomena will also be present for the Critical State Model (CSM), which can be explained by the fact that any $|J|$ such that $|J|\le J_c$ is allowed, instead of only $|J|=J_c$ or 0.

In conclusion, the presented variational method is able to model fully 3D situations for any vector $\vE(\vJ)$ relation, being promising to describe force-free effects or coupling currents in multi-filamentary wires or tapes. The implementation in C++ with OpenMP is time efficient and requires low computer memory. Future work will be dedicated to adapt the parallel routine to the MPI protocol for computer clusters\E{, as well as taking anisotropic $\vE(\vJ)$ relations with force-free effects}.


\section*{Acknowledgements}

We acknowledge valuable discussions with Leonid Prigozhin. The authors acknowledge the use of resources provided by the SIVVP project (ERDF, ITMS 26230120002), the finantial support of the Grant Agency of the Ministry of Education of the Slovak Republic and the Slovak Academy of Sciences (VEGA) under contract no. 2/0126/15\E{, as well as the R\&D Operational Program funded by the ERDF under Grant ITMS 26240120019 `CENTE II'(0.5)}.


\appendix

\section{Variational calculus for functionals with double volume integrals}
\label{s.delL}

This appendix deduces the expressions for the Euler Partial Differential Equations (PDE), the variation, $\delta L$, and second variation, $\delta^2 L$, of functionals containing double volume integrals in multiple dimensions, as those in our variational principle of section \ref{s.varprin}. Although the expression can be deduced from the general mathematical framework of \cite{courant}, these expressions are not present in that book. We also iclude the well-known expressions in \cite{courant} for single integrals, for completeness.


\subsection{Functionals with single volume integrals}
\label{s.Lsingle}

Here we assume a functional with $n$ variables, $\{r_\alpha\}$ with $\alpha\in\{1,\dots,n\}$, and $m$ functions, $\{u_i\}$ with $i\in\{1,\dots,m\}$ and $u_i(\{r_\alpha\})$, where the functional density $f$ also depends on the variable derivative of the functions, $\{u_i^{(\alpha)}\}$, where $u_i^{(\alpha)}\equiv \partial_\alpha u_i\equiv \partial u_i/\partial r_\alpha$, that includes a single $n$-volume integral
\begin{equation}
L[\{u_i\}]=\int_\Omega\dvoln\ f(\{r_\alpha\},\{u_i\},\{u_i^{(\alpha)}\}) ,
\end{equation}
where the integration volume $\Omega$ is made in the whole $n$-space. A finite well-defined value of the functional requires that $f$ vanishes at least at infinity.

For a small change in $\{u_i\}$ proportional to $\epsilon$ as $\{u_i+\epsilon g_i\}$, where $\{g_i\}$ is any set of functions of $\{r_\alpha\}$, the functional can be expanded in a Taylor series up to second order
\begin{equation}
L[\{u_i+\epsilon g_i\}]\approx L[\{u_i\}]+\delta L[\{u_i\}] + \delta^2 L[\{u_i\}], \label{LTaylor}
\end{equation}
where the first, $\delta L[\{u_i\}]$, and second, $\delta^2 L[\{u_i\}]$, variations are defined as
\begin{eqnarray}
&& \delta L[\{u_i\}] \equiv \epsilon \frac{\dif}{\dif\epsilon}\left ( L[\{u_i+\epsilon g_i\}] \right )_{\epsilon=0} \label{dLndef}\\
&& \delta^2 L[\{u_i\}] \equiv \half \epsilon^2 \frac{\dif^2}{\dif\epsilon^2}\left ( L[\{u_i+\epsilon g_i\}] \right )_{\epsilon=0} \label{d2Lndef} .
\end{eqnarray}

The Euler equations are the PDE that follow when the functional is extremal. This occurs when the variation vanishes, $\delta L$=0. Using that $\partial_\alpha(u_i+\epsilon g_i)=u_i^{(\alpha)}+\epsilon g_i^{(\alpha)}$, the variation is
\begin{equation}
\delta L[\{u_i\}]=\epsilon\int_\Omega\dvoln \left ( \frac{\dif }{\dif\epsilon} f(\{r_\alpha\},\{u_i+\epsilon g_i\},\{u_i^{(\alpha)}+\epsilon g_i^{(\alpha)}\}) \right )_{\epsilon=0} ,
\label{dLsin}
\end{equation}
with
\begin{equation}
\left ( \frac{\dif f}{\dif \epsilon} \right )_{\epsilon=0} =\fui g_i+\fuia g_i^{(\alpha)}
\end{equation}
where $f^{(u_i)}\equiv {\partial f}/{\partial u_i}$ and $\fuia\equiv {\partial f}/{\partial u_i^{(\alpha)}}$ and we used Einstein's notation for the summation; for example, $\fuia g_i^{(\alpha)}$ corresponds to $\sum_{i,j=1}^{i=m, j=n}\fuia g_i^{(\alpha)}$. Next, we integrate (\ref{dLsin}) by parts. For this purpose, we use that $\partial_\alpha[\fuia g_i]=\partial_\alpha \fuia g_i+\fuia g_i^{(\alpha)}$ and we apply the generalized divergence theorem for an $n$-vector field with components $\{A_\alpha\}$
\begin{equation}
\int_\Omega\dvoln\partial_\alpha A_\alpha = \int_{\partial\Omega}\dif S_{n-1}n_\alpha A_\alpha , 
\end{equation}
where $\{n_\alpha\}$ are the components of the normal vector to the surface in $n-1$ dimensions, $\partial\Omega$, that encloses the volume $\Omega$ and $\dif S_{n-1}$ is the surface differential. Since $f$ vanishes at the infinite, $\int_{\partial\Omega}\dif S_{n-1}n_\alpha\fuia g_i$ also vanishes, and hence
\begin{equation}
\int_\Omega\dvoln\ \fuia g_i^{(\alpha)} = -\int_\Omega\dvoln\ \partial_\alpha\fuia g_i .
\end{equation}
Then, 
\begin{equation}
\delta L[\{u_i\}]=\epsilon\int_\Omega\dvoln\ g_i\left [ \fui - \partial_\alpha\fuia \right ]
\end{equation}
and the condition $\delta L=0$ follows for any $\{g_i\}$, if and only if
\begin{equation}
\fui-\partial_\alpha\fuia=0 ,
\end{equation}
which are the Euler PDE of the functional.

The condition $\delta L=0$ only imposes that the functional is extremal. As seen from equations (\ref{LTaylor}-\ref{d2Lndef}), the extreme is a minimum when the second variation is positive there, $\delta^2 L>0$, for any $\{g_i\}$. When $\delta^2 L >0$ also applies for any functions $\{u_i\}$, whether they are extremal or not, the minimum is unique. From (\ref{d2Lndef}), $\delta^2L$ is
\begin{equation}
\delta^2 L[\{u_i\}]=\half\epsilon^2\int_\Omega\dvoln \left ( \frac{\dif^2 }{\dif\epsilon^2} f(\{r_\alpha\},\{u_i+\epsilon g_i\},\{u_i^{(\alpha)}+\epsilon g_i^{(\alpha)}\}) \right )_{\epsilon = 0}
\label{d2L}
\end{equation}
with 
\begin{equation}
\left ( \frac{\dif^2f}{\dif\epsilon^2} \right )_{\epsilon=0} =\fuij g_ig_j+\fuiajb g_i^{(\alpha)}g_j^{(\beta)} + 2 \fuijb g_ig_j^{(\beta)} ,
\label{d2LdV1}
\end{equation}
where $\fuij\equiv \partial^2 f/(\partial u_i\partial u_j)$, and similarly for $\fuiajb$ and $\fuijb$.


\subsection{Functionals with double volume integrals}

Many phenomena in physics require variational principles from functionals containing double integrals of the $n$-volume, as those in this article. The general form of these functionals is
\begin{equation}
L[\{u_i\}]=\int_\Omega\dvoln\int_\Omega\dvoln'\ f(\{r_\alpha\},\{r_\alpha'\},\{u_i\},\{u_i'\},\{u_i^{(\alpha)}\},\{{u_i'}^{(\alpha)}\}) ,
\end{equation}
where $u_i'$ is the same function as $u_i$ but with variables $\{r_\alpha'\}$ instead of $\{r_\alpha\}$, $u_i'=u_i(\{r_\alpha'\})$, and ${u_i'}^{(\alpha)}\equiv\partial_\alpha'u_i'\equiv\partial u_i'/\partial r_\alpha'$. 

Similar to functionals with single volume integrals, the variation from (\ref{dLndef}) becomes
\begin{eqnarray}
\delta L[\{u_i\}] & = & \epsilon\int_\Omega\dvoln\int_\Omega\dvoln' \Bigg( \frac{\dif }{\dif\epsilon} f(\{r_\alpha\},\{r_\alpha'\},\{u_i+\epsilon g_i\},\{u_i'+\epsilon g_i'\}, \nonumber \\
&&  \{u_i^{(\alpha)}+\epsilon g_i^{(\alpha)}\},\{{u_i'}^{(\alpha)}+\epsilon {g_i'}^{(\alpha)}\} ) \Bigg)_{\epsilon=0} ,
\end{eqnarray}
with
\begin{equation}
\left ( \frac{\dif f}{\dif \epsilon} \right )_{\epsilon=0}=\fui g_i + \fuia g_i^{(\alpha)} + \fuip g_i' + \fuiap {g_i'}^{(\alpha)}.
\end{equation}
After integrating by parts,
\begin{eqnarray}
\delta L[\{u_i\}] & = & \epsilon\int_\Omega\dvoln\int_\Omega\dvoln' \left [ g_i \left ( \fui - \partial_\alpha\fuia \right ) + \right . \nonumber\\
&& \left . g_i'\left ( \fuip-\partial_\alpha'\fuiap \right ) \right ] .
\end{eqnarray}
In physics, the functional density is usually symmetric with respect to $\{r_\alpha\}$ and $\{r_\alpha'\}$ and respective functions, as
\begin{eqnarray}
&& f(\{r_\alpha\},\{r_\alpha'\},\{u_i\},\{u_i'\},\{u_i^{(\alpha)}\},\{{u_i'}^{(\alpha)}\}) 
= \nonumber \\ && 
f(\{r_\alpha'\},\{r_\alpha\},\{u_i'\},\{u_i\},\{{u_i'}^{(\alpha)}\},\{{u_i}^{(\alpha)}\}) , 
\label{Lsym'}
\end{eqnarray}
which is the case of our functionals in (\ref{LJ}), (\ref{LTint}) and (\ref{LTB}). Applying this symmetry,
\begin{equation}
\delta L[\{u_i\}] = 2\epsilon\int_\Omega\dvoln g_i \int_\Omega\dvoln' \left [ \fui - \partial_\alpha\fuia \right ]
\end{equation}
and the Euler PDE corresponding to $\delta L=0$ for any arbitrary function $g_i$ are
\begin{equation}
2\int_\Omega\dvoln' \left [ \fui - \partial\alpha\fuia \right ]=0.
\end{equation}
For functionals with a combination of single and double integrals as
\begin{eqnarray}
L[\{u_i\}] & = & \int_\Omega\dvoln h(\{r_\alpha\},\{u_i\},\{u_i^{(\alpha)}\}) \nonumber\\
& + & \int_\Omega\dvoln\int_\Omega\dvoln'\ f(\{r_\alpha\},\{r_\alpha'\},\{u_i\},\{u_i'\},\{u_i^{(\alpha)}\},\{{u_i'}^{(\alpha)}\}) ,
\label{Lhf}
\end{eqnarray}
we only need to add both contributions to $\delta L$, obtaining
\begin{eqnarray}
\delta L[\{u_i\}] & = & \epsilon\int_\Omega\dvoln g_i \Bigg[ \hui - \partial_\alpha\huia \nonumber\\
& + & \left . 2\int_\Omega\dvoln' \left ( \fui - \partial_\alpha\fuia \right ) \right ]
\label{dLboth}
\end{eqnarray}
and the corresponding Euler PDE
\begin{eqnarray}
 \hui - \partial_\alpha\huia + \left [ 2\int_\Omega\dvoln' \left ( \fui - \partial_\alpha\fuia \right ) \right ] = 0.
\label{EulerdV2}
\end{eqnarray}

The second variation from (\ref{d2Lndef}) is 
\begin{eqnarray}
\delta^2 L[\{u_i\}] & = & \half\epsilon^2\int_\Omega\dvoln\int_\Omega\dvoln' \Bigg( \frac{\dif^2 }{\dif\epsilon^2} f(\{r_\alpha\},\{r_\alpha'\},\{u_i+\epsilon g_i\},\{u_i'+\epsilon g_i'\}, \nonumber \\
&&  \{u_i^{(\alpha)}+\epsilon g_i^{(\alpha)}\},\{{u_i'}^{(\alpha)}+\epsilon {g_i'}^{(\alpha)}\} ) \Bigg)_{\epsilon=0} ,
\end{eqnarray}
with
\begin{eqnarray}
\left ( \frac{\dif^2 f}{\dif \epsilon^2} \right )_{\epsilon=0} & = & \fuij g_ig_j + f^{(u_i'u_j')}g_i'g_j' + 2\fuijb g_ig_i^{(\beta)} +
2f^{(u_i'{u_j'}^{(\beta)})} g_i'{g_j'}^{(\beta)} \nonumber\\
& + & \fuiajb g_i^{(\alpha)}g_j^{(\beta)} + f^{( {u_i'}^{(\alpha)}{u_j'}^{(\beta)} )} {g_i'}^{(\alpha)}{g_j'}^{(\beta)} \nonumber\\
& + & 2f^{( {u_i}{u_j'}^{(\beta)} )}g_i{g_j'}^{(\beta)} + 2f^{( {u_i'}{u_j}^{(\beta)} )}{g_i'}{g_j}^{(\beta)} \nonumber \\
& + & 2f^{( {u_i}{u_j'} )}g_i{g_j'} + 2f^{( {u_i}^{(\alpha)}{u_j'}^{(\beta)} )} {g_i}^{(\alpha)}{g_j'}^{(\beta)},
\end{eqnarray}
where $f^{(u_i{u_j'}^{(\beta)})}\equiv \partial^2f/(\partial u_i\partial {u_j'}^{(\beta)})$, and similarly with all the other terms. If the functional density follows the symmetry of (\ref{Lsym'}), the second variation is simplified as
\begin{eqnarray}
\delta^2 L[\{u_i\}] & = & \half\epsilon^2\int_\Omega\dvoln\int_\Omega\dvoln' \left [ 2\fuij g_ig_j + 2\fuiajb g_i^{(\alpha)}g_j^{(\beta)} \right . \nonumber \\
& + & 2f^{( {u_i}{u_j'} )}g_i{g_j'} + 2f^{( {u_i}^{(\alpha)}{u_j'}^{(\beta)} )} {g_i}^{(\alpha)}{g_j'}^{(\beta)} \nonumber\\
& + & \left . 4\fuijb g_ig_i^{(\beta)} + 4f^{( {u_i}{u_j'}^{(\beta)} )}g_i{g_j'}^{(\beta)} \right ] .
\label{d2LdV2}
\end{eqnarray}
When the functional contains both terms with single and double volume integrals like (\ref{Lhf}), the second variation is
\begin{eqnarray}
\delta^2 L[\{u_i\}] & = & \half\epsilon^2\int_\Omega\dvoln \nonumber \left [ h^{(u_iu_j)}g_ig_j + h^{(u_i^{(\alpha)}u_j^{(\beta)})}g_i^{(\alpha)}g_j^{(\beta)} + 2h^{(u_iu_j^{(\beta)})}g_ig_j^{(\beta)} \right ] \\
& + & \half\epsilon^2\int_\Omega\dvoln\int_\Omega\dvoln' \left [ 2\fuij g_ig_j + 2\fuiajb g_i^{(\alpha)}g_j^{(\beta)} \right . \nonumber \\
& + & 2f^{( {u_i}{u_j'} )}g_i{g_j'} + 2f^{( {u_i}^{(\alpha)}{u_j'}^{(\beta)} )} {g_i}^{(\alpha)}{g_j'}^{(\beta)} \nonumber\\
& + & \left . 4\fuijb g_ig_i^{(\beta)} + 4f^{( {u_i}{u_j'}^{(\beta)} )}g_i{g_j'}^{(\beta)} \right ] .
\label{d2Lboth}
\end{eqnarray}



\section{Evaluation of variables for the discretized problem}
\label{s.vardis}

This appendix contains details and the formulas to calculate relevant quantities, such as $\vJ$, $\vA$ and the functional, from our discretization of $\vT$ in section \ref{s.dis}.

For our discretization, the current density $\vJ$ at any point $\vr$ can be found by linear interpolation as,
\begin{equation}
J_s(\vr)=\sum_{i=1}^{n_s} J_{si}h_{si}(\vr)
\label{Jinter}
\end{equation}
where $s\in\{x,y,z\}$, $J_s$ is the $s$ component of $\vJ$, $n_s$ is the number of surfaces perpendicular to the $s$ direction (or $s$-surfaces), $J_{si}$ is $J_s$ at surface $i$, and $h_{si}(\vr)$ is the interpolation function that decreases linearly in the $s$ direction from 1 at surface $i$ to 0 at the neighboring surfaces and vanishes elsewhere. Since, $\vJ=\rotT$, $J_{si}$ in one $s$-surface is related to $\vT$ as
\begin{equation}
J_{si}=\frac{1}{S_{si}}\int_{S_{si}}\dsur\cdot(\rotT)=\frac{1}{S_{si}}\int_{\partial S_{si}}\dif{\bf l}\cdot\vT ,
\end{equation}
where $S_i$ is the area of the $s$-surface $i$, $\partial S_{si}$ is the contour of that surface, and $\dsur$ and $\dif{\bf l}$ are the surface and line differentials, respectively. For our discretization, the equation above results in
\begin{equation}
J_{si}=\sum_{q\in\{x,yz\}}\sum_{j=1}^{m_q} R_{sqij}T_{qj} ,
\label{JsiTqj}
\end{equation}
where $m_q$ are the number of edges parallel to the $q$ direction and the matrix with elements $R_{sqij}$ is sparsely filled, being non-zero only for the edges in the contour of the surface with indexes $s,i$.

In order to discretize the functional (\ref{LTint}), we write it as a function of $\vJ$ and take into account that for the decomposition of (\ref{Jinter}), 
\begin{equation}
\frac{\mu_0}{4\pi}\int_V\dif V\int_V\dif V'\frac{\vJ(\vr)\cdot\vJ(\vr')}{|\vr-\vr'|}=\sum_{s\in\{x,y,z\}}\sum_{i,j=1}^{n_s}V_{si}V_{sj}J_{si}J_{sj}a_{sij}
\label{intdis}
\end{equation}
where the interaction matrix elements $a_{sij}$ are
\begin{equation}
a_{sij}\equiv \frac{\mu_0}{4\pi V_{si}V_{sj}} \int\dif V\int\dif V' \frac{h_{si}(\vr)h_{sj}(\vr')}{|\vr-\vr'|}
\label{adef}
\end{equation}
with
\begin{equation}
V_{si}\equiv  \int\dif V h_{si}(\vr) .
\end{equation}
In consistence with (\ref{intdis}), the vector potential at the $s$-surface is defined as 
\begin{equation}
A_{si}=\int\dif V A_s(\vr)h_{si}(\vr)=\sum_{j=1}^{n_s}V_{sj}J_{sj}a_{sij} .
\end{equation}
Taking this into account, the functional in (\ref{LTint}) becomes
\begin{eqnarray}
L & = & \frac{1}{2\Delta t}\sum_{s\in\{x,y,z\}}\sum_{i,j=1}^{n_s}V_{si}V_{sj}\Delta J_{si}\Delta J_{sj}a_{sij} \nonumber \\
& + & \sum_{s\in\{x,y,z\}}\sum_{i=1}^{n_s}V_{si}\Delta J_{si}\Delta A_{a,si} + \sum_{\alpha=1}^{N}V_{\alpha}U_{\alpha} ,
\end{eqnarray}
where $V_{\alpha}$ is the volume of cell $\alpha$, $N$ is the total number of cells, $U_{\alpha}$ is defined as $U_{\alpha}\equiv U(\vJ(r_{\alpha}))$, $\vr_{\alpha}$ is the center of cell $\alpha$, $\vJ(\vr_{\alpha})$ is the interpolated current density obtained by (\ref{Jinter}), and $\Delta A_{a,si}$ is
\begin{equation}
\Delta A_{a,si}=\int \dif V\ \Delta A_{a,s}(\vr)h_{si}(\vr).
\end{equation}
Using (\ref{JsiTqj}), the functional as a function of the $\Delta \vT$ at the edges is
\begin{eqnarray}
L & = & \sum_{s,p,q\in\{x,y,z\}} \sum_{i,k=1}^{n_s} \sum_{j=1}^{m_p} \sum_{k=1}^{m_q} \frac{1}{2\Delta t}V_{si}V_{sk}a_{sik}R_{spij}R_{sqkl}\Delta T_{pj}\Delta T_{ql} \nonumber \\
& + & \sum_{s,p\in\{x,y,z\}}\sum_{i=1}^{n_s}\sum_{j=1}^{m_p} \frac{1}{\Delta t}\Delta A_{a,si}V_{si}R_{spij}\Delta T_{pj} + \sum_{\alpha=1}^N V_{\alpha}U_{\alpha}.
\label{LTdis}
\end{eqnarray}

\begin{figure}[ptb]
\centering
{\includegraphics[trim=0 0 0 0,clip,width=5.0 cm]{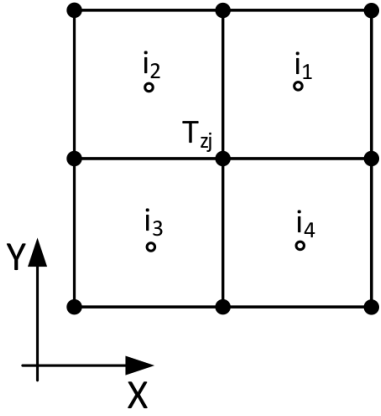}}
\caption{Definition of cell indexes $i_1,i_2,i_3,i_4$ neighboring edge $zj$ in (\ref{czj}).}
\label{f.c}
\end{figure}

From (\ref{LTdis}) we can find that the change in the functional due to a change $v$ at the $p$-edge $j$, so that $\Delta T_{pj}:=\Delta T_{pj}+v$, is
\begin{eqnarray}
dL_{pj} & = & \frac{1}{\Delta t}\Delta F_{pj}v
+ \frac{1}{2\Delta t}G_{pj}v^2 \nonumber \\
& + & \sum_{\alpha\in N_{pj}} V_{\alpha}[ U(\vJ(\vr_{\alpha})+{\bf c}_{\alpha pj}v)-U_\alpha ]
\end{eqnarray}
with
\begin{eqnarray}
\Delta F_{pj}=\sum_{s\in\{x,y,z\}}\sum_{i=1}^{n_{spj}} (\Delta A_{J,si} + \Delta A_{a,si})V_{si}R_{spij}, \label{Fpj} \\
G_{pj}=\sum_{s\in\{x,y,z\}}\sum_{i,k=1}^{n_{spj}} V_{si}V_{sk}a_{sik}R_{spij}R_{spkj}.
\end{eqnarray}
In the equations above, $n_{spj}$ is the set of $s$-surfaces meeting at edge $pj$; $\Delta A_{J,si}$ is the $s$ component of the vector potential at the $s$-surface $i$ created by $\Delta \vJ$, 
\begin{equation}
\Delta A_{J,si}=\sum_{j=1}^{n_s}V_{sj}\Delta J_{sj}a_{sij} ; 
\end{equation}
constant $G_{pj}$ is the self-interaction term; $N_{pj}$ is the set of cells neighboring edge $pj$, and ${\bf c}_{\alpha pj}v$ is the change of interpolated $\vJ$ at $\vr_{\alpha}$ due to the change $v$ in $\Delta T_{pj}$. For example, in uniform mesh and edges following the $z$ axis, so that $p=z$, ${\bf c}_{\alpha zj}$ takes the form
\begin{eqnarray}
{\bf c}_{i_1zj}=\half\left ( -\frac{1}{l_y}{\bf e}_x + \frac{1}{l_x}{\bf e}_y \right ) \nonumber \\
{\bf c}_{i_2zj}=\half\left ( -\frac{1}{l_y}{\bf e}_x - \frac{1}{l_x}{\bf e}_y \right ) \nonumber \\
{\bf c}_{i_3zj}=\half\left ( \frac{1}{l_y}{\bf e}_x - \frac{1}{l_x}{\bf e}_y \right ) \nonumber \\
{\bf c}_{i_4zj}=\half\left ( \frac{1}{l_y}{\bf e}_x + \frac{1}{l_x}{\bf e}_y \right ) ,
\label{czj}
\end{eqnarray} 
where the cell indexes $i_1,i_2,i_3,i_4$ relative to edge $zj$ are defined as in figure \ref{f.c}, $l_x$ and $l_y$ are the cells size in the $x$ and $y$ directions, respectively, and ${\bf e}_x$ and ${\bf e}_y$ are the unit vectors in the $x$ and $y$ directions, respectively. Quantity $\Delta F_{pj}$ in (\ref{Fpj}) is proportional to the magnetic flux density $\vB=\nabla\times\vA$ at edge $pj$.

In order to evaluate the dissipation function for a $\vB$-dependent critical current density [or any $\vE(\vJ,\vB)$], we compute the average magnetic flux density at any cell $\alpha$ created by our discretized $\vJ$ in (\ref{Jinter}), resulting in
\begin{equation}
\vB_{\alpha}\equiv\frac{1}{V_\alpha}\int_{V_\alpha}\dif V\ \vB(\vr)=\sum_{s\in\{x,y,z\}}\sum_{j=1}^{n_s}V_{sj}J_{sj}{\bf e}_s\times{\bf d}_{\alpha sj}
\end{equation}
with
\begin{equation}
{\bf d}_{\alpha sj}\equiv \frac{\mu_0}{4\pi V_\alpha V_{sj}}\int_{V_j}\dif V\int\dif V' \frac{ h_{sj}(\vr')(\vr-\vr') }{|\vr-\vr'|^3},
\label{ddef}
\end{equation}
where ${\bf e}_s$ is the unit vector in the direction of axis $s$.

\E{
For uniform rectangular mesh, the interaction matrix of the vector potential in (\ref{adef}), $a_{sij}$, can be drastically reduced. For this case, we can label each $s$-surface by three independent indexes ($i_x$,$i_y$,$i_z$) instead of a global index $i$. Then, the interaction matrix elements $a_{sij}$ can be labeled as $a_{si_xi_yi_zj_xj_yj_z}$. Thanks to the regular rectangular mesh, the interaction matrix obeys the discrete translation symmetry
\begin{equation}
a_{si_xi_yi_zj_xj_yj_z}=a_{s111|j_x-i_x+1||j_y-i_y+1||j_z-i_z+1|}\equiv a_{s111k_xk_yk_z}.
\end{equation}
Since $k_x,k_y,k_z$ are between 1 and $n_{sx}$,$n_{sy}$,$n_{sz}$, respectively, being the latter the number of $s$-surfaces in the $x,y,z$ directions, respectively, there are only $n_{sx}n_{sy}n_{sz}$ independent matrix entries, while in the complete matrix for the $s$-surfaces thre are as many as $(n_{sx}n_{sy}n_{sz})^2$ entries. Then, for an object with $n_x\times n_y\times n_z$ cells in the $x,y,z$ directions, respectively, we reduce the total interaction matrix from $[(n_x+1)n_yn_z]^2+[n_x(n_y+1)n_z]^2+[n_xn_y(n_z+1)]^2$ entries to only $(n_x+1)n_yn_z+n_x(n_y+1)n_z+n_xn_y(n_z+1)$, being the total number of surfaces. For a cube of $n_x=n_y=n_z=41$,} we reduce the RAM memory from around 117 Gb (estimated) to 1.7 Mb. A similar reduction can be achieved with the interaction matrices for the magnetic field. \E{For this kind of reduction, the cells do not need to be necessarily cubic.}

In this work, we approximate the interaction matrices $a_{sij}$ and ${\bf d}_{\alpha sj}$ in (\ref{adef}) and (\ref{ddef}), respectively, as follows. The matrix elements $a_{sij}$ are
\begin{eqnarray}
a_{sij} & \approx & \frac{\mu_0}{4\pi|\vr_{si}-\vr_{sj}|}\qquad \textrm{if $i\neq j$} \nonumber\\
& \approx & \frac{\mu_0}{4\pi V_{si}^2} \int_{V_{si}}\dif V\int_{V_{si}}\dif V' \frac{1}{|\vr-\vr'|} \qquad \textrm{if $i=j$} ,
\end{eqnarray}
where $\vr_{si}$ and $\vr_{sj}$ are the center position of surfaces $si$ and $sj$, respectively, and the volume integral expands over the rectangular prism with base $S_{si}$ and height corresponding to the segment in the $s$ direction joining the center of neighboring cells. This integral is analytical \E{for any rectangular prism}, although we do not include the expression here for space reasons. \E{The expression for a cube is very simple, which can be found from that of a uniformly charged cube \cite{ciftja11PLA} as $a_{sii}\approx\mu_0/(4\pi L_i)\{(1+\sqrt{2}-2\sqrt{3})/5-\pi/3+\ln [(1+\sqrt{2})(2+\sqrt{3})]\}$, where $L_i$ is the side of the cube associated to surface $i$.} For ${\bf d}_{\alpha sj}$, we take
\begin{equation}
{\bf d}_{\alpha sj}\approx \frac{\mu_0(\vr_\alpha-\vr_{sj})}{4\pi |\vr_\alpha-\vr_{sj}|^3}.
\end{equation}
This approach is effective for cubic mesh in 3D or square mesh for thin films, as that in this article. For elements elongated in one direction, these matrices are integrated numerically by dividing each element into smaller sub-elements and using the equations above for the sub-elements.


\section*{References}

\begin{thebibliography}{10}
\expandafter\ifx\csname url\endcsname\relax
  \def\url#1{\texttt{#1}}\fi
\expandafter\ifx\csname urlprefix\endcsname\relax\def\urlprefix{URL }\fi
\expandafter\ifx\csname href\endcsname\relax
  \def\href#1#2{#2} \def\path#1{#1}\fi

\bibitem{acreview}
F.~Grilli, E.~Pardo, A.~Stenvall, D.~N. Nguyen, W.~Yuan, F.~G{\"om\"o}ry,
  Computation of losses in {HTS} under the action of varying magnetic fields
  and currents, IEEE Trans. Appl. Supercond. 24~(1) (2014) 8200433.

\bibitem{vlaskovlasov15PRB}
V.~Vlasko-Vlasov, A.~Koshelev, A.~Glatz, C.~Phillips, U.~Welp, W.~Kwok, Flux
  cutting in high-$t_c$ superconductors, Phys. Rev. B 91~(1) (2015) 014516.

\bibitem{mishev15SST}
V.~Mishev, M.~Zehetmayer, D.~Fischer, M.~Nakajima, H.~Eisaki, M.~Eisterer,
  Interaction of vortices in anisotropic superconductors with isotropic
  defects, Supercond. Sci. Technol. 28~(10) (2015) 102001.

\bibitem{pecher04ICS}
R.~Pecher, M.~McCulloch, S.~Chapman, L.~Prigozhin, C.~Elliott, {3D}-modelling
  of bulk type{-II} superconductors using unconstrained {H-}formulation, Inst.
  of Phys.: Conf. Ser. 181 (2003) 1418, european Conference on Applied
  Superconductivity ({EUCAS}) 2003.

\bibitem{zehetmayer06SST}
M.~Zehetmayer, M.~Eisterer, H.~Weber, Simulation of the current dynamics in a
  superconductor induced by a small permanent magnet: application to the
  magnetoscan technique, Supercond. Sci. Technol. 19 (2006) S429.

\bibitem{zhangM12SSTa}
M.~Zhang, T.~Coombs, {3D} modeling of high{-$T_c$} superconductors by finite
  element software, Supercond. Sci. Technol. 25 (2012) 015009.

\bibitem{grilli13Cry}
F.~Grilli, R.~Brambilla, F.~Sirois, A.~Stenvall, S.~Memiaghe, Development of a
  three-dimensional finite-element model for high-temperature superconductors
  based on the {$H$}-formulation, Cryogenics 53 (2013) 142--147.

\bibitem{zermeno14SSTa}
V.~M.~R. Zermeno, F.~Grilli, {3D} modeling and simulation of {2G HTS} stacks
  and coils, Supercond. Sci. Technol. 27 (2014) 044025.

\bibitem{stenvall14SST}
A.~Stenvall, V.~Lahtinen, M.~Lyly, An {H-formulation-based} three-dimensional
  hysteresis loss modelling tool in a simulation including time varying applied
  field and transport current: the fundamental problem and its solution,
  Supercond. Sci. Technol. 27~(10) (2014) 104004.

\bibitem{escamez16IES}
G.~Escamez, F.~Sirois, V.~Lahtinen, A.~Stenvall, A.~Badel, P.~Tixador,
  B.~Ramdane, G.~Meunier, R.~Perrin-Bit, C.-E. Bruzek, {3-D} numerical modeling
  of {AC} losses in multifilamentary mgb$_2$ wires, IEEE Trans. Appl.
  Supercond. 26~(3) (2016) 1--7.

\bibitem{grilli05IES}
F.~Grilli, S.~Stavrev, Y.~Le~Floch, M.~Costa-Bouzo, E.~Vinot, I.~Klutsch,
  G.~Meunier, P.~Tixador, B.~Dutoit, Finite-element method modeling of
  superconductors: from {2-D} to {3-D}, IEEE Trans. Appl. Supercond. 15~(1)
  (2005) 17--25.

\bibitem{lousberg09SST}
G.~Lousberg, M.~Ausloos, C.~Geuzaine, P.~Dular, P.~Vanderbemden,
  B.~Vanderheyden, Numerical simulation of the magnetization of
  high-temperature superconductors: a {3D} finite element method using a single
  time-step iteration, Supercond. Sci. Technol. 22 (2009) 055005.

\bibitem{fagnard16SST}
J.-F. Fagnard, M.~Morita, S.~Nariki, H.~Teshima, H.~Caps, B.~Vanderheyden,
  P.~Vanderbemden, Magnetic moment and local magnetic induction of
  superconducting/ferromagnetic structures subjected to crossed fields:
  experiments on {GdBCO} and modelling, Supercond. Sci. Technol. 29~(12) (2016)
  125004.

\bibitem{campbell09SST}
A.~M. Campbell, A direct method for obtaining the critical state in two and
  three dimensions, Supercond. Sci. Technol. 22 (2009) 034005.

\bibitem{komi09PhC}
Y.~Komi, M.~Sekino, H.~Ohsaki, Three-dimensional numerical analysis of magnetic
  and thermal fields during pulsed field magnetization of bulk superconductors
  with inhomogeneous superconducting properties, Physica C 469~(15) (2009)
  1262--1265.

\bibitem{farinon14SST}
S.~Farinon, G.~Iannone, P.~Fabbricatore, U.~Gambardella, {2D} and {3D}
  numerical modeling of experimental magnetization cycles in disks and spheres,
  Supercond. Sci. Technol. 27~(10) (2014) 104005.

\bibitem{prigozhin96JCP}
L.~Prigozhin, The bean model in superconductivity: Variational formulation and
  numerical solution, J. Comput. Phys. 129~(1) (1996) 190--200.

\bibitem{prigozhin97IES}
L.~Prigozhin, Analysis of critical-state problems in type{-II}
  superconductivity, IEEE Trans. Appl. Supercond. 7~(4) (1997) 3866--3873.

\bibitem{prigozhin98JCP}
L.~Prigozhin, Solution of thin film magnetization problems in {type-II}
  superconductivity, J. Comput. Phys. 144~(1) (1998) 180--193.

\bibitem{prigozhin11SST}
L.~Prigozhin, V.~Sokolovsky, Computing {AC} losses in stacks of
  high-temperature superconducting tapes, Supercond. Sci. Technol. 24 (2011)
  075012.

\bibitem{HacIacinphase}
E.~Pardo, F.~G{\"o}m{\"o}ry, J.~{\v{S}}ouc, J.~Ceballos, Current distribution
  and ac loss for a superconducting rectangular strip with in-phase alternating
  current and applied field, Supercond. Sci. Technol. 20~(4) (2007) 351--364.

\bibitem{pancaketheo}
E.~Pardo, Modeling of coated conductor pancake coils with a large number of
  turns, Supercond. Sci. Technol. 21 (2008) 065014.

\bibitem{pardo15SST}
E.~Pardo, J.~{\v Souc}, L.~{Frolek}, Electromagnetic modelling of
  superconductors with a smooth current-voltage relation: variational principle
  and coils from a few turns to large magnets, Supercond. Sci. Technol. 28
  (2015) 044003.

\bibitem{sanchez06JAP}
A.~Sanchez, N.~Del~Valle, E.~Pardo, D.-X. Chen, C.~Navau, Magnetic levitation
  of superconducting bars, J. Appl. Phys. 99~(11) (2006) 113904.

\bibitem{via15SST}
G.~Via, N.~Del-Valle, A.~Sanchez, C.~Navau, Simultaneous magnetic and transport
  currents in thin film superconductors within the critical-state
  approximation, Supercond. Sci. Technol. 28~(1) (2015) 014003.

\bibitem{ruuskanen14IES}
J.~Ruuskanen, A.~Stenvall, V.~Lahtinen, Utilizing triangular mesh with {MMEV}
  to study hysteresis losses of round superconductors obeying critical state
  model, IEEE Trans. Appl. Supercond. 25~(3) (2014) 8200405,
  10.1109/TASC.2014.2365408.

\bibitem{zhangY15SST}
Y.~Zhang, Y.~Song, L.~Wang, X.~Liu, Simulation of superconducting tapes and
  coils with convex quadratic programming method, Supercond. Sci. Technol.
  28~(8) (2015) 085002.

\bibitem{brandt95PRL}
E.~Brandt, Square and rectangular thin superconductors in a transverse magnetic
  field, Phys. Rev. Lett. 74~(15) (1995) 3025--3028.

\bibitem{brandt95PRBa}
E.~Brandt, Electric field in superconductors with rectangular cross section,
  Phys. Rev. B 52~(21) (1995) 15442.

\bibitem{brandt96PRB}
E.~H. Brandt, Superconductors of finite thickness in a perpendicular magnetic
  field: {Strips} and slabs, Phys. Rev. B 54~(6) (1996) 4246.

\bibitem{rhyner98PhC}
J.~Rhyner, Calculation of {AC} losses in {HTSC} wires with arbitrary current
  voltage characteristics, Physica C 310~(1-4) (1998) 42--47.

\bibitem{costa04SST}
M.~Costa~Bouzo, F.~Grilli, Y.~Yang, Modelling of coupling between
  superconductors of finite length using an integral formulation, Supercond.
  Sci. Technol. 17~(10) (2004) 1103.

\bibitem{morandi15SST}
A.~Morandi, M.~Fabbri, A unified approach to the power law and the critical
  state modeling of superconductors in {2D}, Supercond. Sci. Technol. 28~(2)
  (2015) 024004.

\bibitem{vestgarden08PRB}
J.~Vestg{\aa}rden, D.~Shantsev, Y.~Galperin, T.~Johansen, Flux distribution in
  superconducting films with holes, Phys. Rev. B 77~(1) (2008) 014521.

\bibitem{vannugteren16IES}
J.~van Nugteren, B.~van Nugteren, P.~Gao, L.~Bottura, M.~Dhall{\'e},
  W.~Goldacker, A.~Kario, H.~ten Kate, G.~Kirby, E.~Krooshoop, et~al.,
  Measurement and numerical evaluation of {AC} losses in a {ReBCO} {Roebel}
  cable at 4.5 k, IEEE Trans. Appl. Supercond. 26~(3) (2016) 1--7.

\bibitem{russenschuck99rep}
S.~Russenschuck (Ed.), {1st} International Roxie Users Meeting and Workshop
  {ROXIE}: routine for the optimization of magnet {X}-sections, inverse field
  calculation and coil end design, CERN, Gen\`eve, 1999.

\bibitem{kurz02IES}
S.~Kurz, S.~Russenschuck, Numerical simulation of superconducting accelerator
  magnets, IEEE Trans. Appl. Supercond. 12~(1) (2002) 1442--1447.

\bibitem{amemiya16SST}
N.~Amemiya, Y.~Sogabe, M.~Sakashita, Y.~Iwata, K.~Noda, T.~Ogitsu, Y.~Ishii,
  T.~Kurusu, Magnetisation and field quality of a cosine-theta dipole magnet
  wound with coated conductors for rotating gantry for hadron cancer therapy,
  Supercond. Sci. Technol. 29~(2) (2016) 024006.

\bibitem{amemiya06JAP}
N.~Amemiya, S.~Sato, T.~Ito, Magnetic flux penetration into twisted
  multifilamentary coated superconductors subjected to ac transverse magnetic
  fields, J. Appl. Phys. 100~(12) (2006) 123907--123907.

\bibitem{nii12SST}
M.~Nii, N.~Amemiya, T.~Nakamura, Three-dimensional model for numerical
  electromagnetic field analyses of coated superconductors and its application
  to roebel cables, Supercond. Sci. Technol. 25~(9) (2012) 095011.

\bibitem{ueda13IES}
H.~Ueda, M.~Fukuda, K.~Hatanaka, T.~Wang, A.~Ishiyama, S.~Noguchi, Spatial and
  temporal behavior of magnetic field distribution due to shielding current in
  {HTS} coil for cyclotron application, IEEE Trans. Appl. Supercond. 23~(3)
  (2013) 4100805--4100805.

\bibitem{magnet10k}
E.~Pardo, Modeling of screening currents in coated conductor magnets containing
  up to 40000 turns, Supercond. Sci. Technol. 29~(8) (2016) 085004.

\bibitem{bossavit94IEM}
A.~Bossavit, Numerical modelling of superconductors in three dimensions: a
  model and a finite element method, IEEE Trans. Magn. 30~(5) (1994)
  3363--3366.

\bibitem{elliott06JNA}
C.~M. Elliott, Y.~Kashima, A finite-element analysis of critical-state models
  for type-{II} superconductivity in {3D}, IMA journal of numerical analysis 27
  (2006) 293--331.

\bibitem{kashima08MNA}
Y.~Kashima, On the double critical-state model for type-{II} superconductivity
  in {3D}, ESAIM: Mathematical Modelling and Numerical Analysis 42~(3) (2008)
  333--374.

\bibitem{badia01PRL}
A.~Bad{\'\i}a, C.~L{\'o}pez, Critical state theory for nonparallel flux line
  lattices in {type-II} superconductors, Phys. Rev. Lett. 87~(12) (2001)
  127004.

\bibitem{badia12SST}
A.~Bad{\'\i}a-Maj{\'o}s, C.~L{\'o}pez, Electromagnetics close beyond the
  critical state: thermodynamic prospect, Supercond. Sci. Technol. 25~(10)
  (2012) 104004.

\bibitem{sanchez01PRB}
A.~Sanchez, C.~Navau, Magnetic properties of finite superconducting cylinders.
  {I.} uniform applied field, Phys. Rev. B 64 (2001) 214506.

\bibitem{tranarr}
E.~Pardo, A.~Sanchez, D.-X. Chen, C.~Navau, Theoretical analysis of the
  transport critical-state ac loss in arrays of superconducting rectangular
  strips, Phys. Rev. B 71 (2005) 134517.

\bibitem{couplingEUCAS}
E.~Pardo, M.~Kapolka, J.~Kov{\'a}{\v{c}}, J.~{\v{S}}ouc, F.~Grilli,
  A.~Piqu{\'e}, Three-dimensional modeling and measurement of coupling {AC}
  loss in soldered tapes and striated coated conductors, IEEE Trans. Appl.
  Supercond. 26~(3) (2016) 1--7.

\bibitem{chen89JAP}
D.-X. Chen, R.~B. Goldfarb, Kim model for magnetization of type-{II}
  superconductors, J. Appl. Phys. 66~(6) (1989) 2489--2500.

\bibitem{navau08JAP}
C.~Navau, A.~Sanchez, N.~Del-Valle, D.~X. Chen, Alternating current
  susceptibility calculations for thin-film superconductors with regions of
  different critical-current densities, J. Appl. Phys. 103 (2008) 113907.

\bibitem{badia05APL}
A.~Bad{\'\i}a-Maj{\'o}s, C.~L{\'o}pez, Critical state model in superconducting
  parallelepipeds, Appl. Phys. Lett. 86~(20) (2005) 202510.

\bibitem{badia15SST}
A.~Bad{\'\i}a-Maj{\'o}s, C.~L{\'o}pez, Modelling current voltage
  characteristics of practical superconductors, Supercond. Sci. Technol. 28~(2)
  (2015) 024003.

\bibitem{kapolka15EUCAS}
M.~Kapolka, E.~Pardo, J.~{Kov\'a\v c}, J.~{\v Souc}, F.~Grilli, R.~Nast,
  E.~{Demenc\'\i k}, A.~{Piqu\'e}, {3D} modeling and measurement of coupling
  {AC} loss in soldered tapes and striated coated conductors, 12th European
  Conference on Applied SuperconductivityPresentation number 3A-LS-O1.8.
  Available at \url{http://snf.ieeecsc.org/file/6056/download?token=qAk3YNJy}.

\bibitem{pardo16HTSmod}
E.~Pardo, M.~Kapolka, Modeling of superconductors interacting with non-linear
  magnetic materials: {3D} variational principles, force-free effects and
  applications, 5th Internatinal Workshop on Numerical Modelling of High
  Temperature Superconductors{DOI:} 10.5281/zenodo.56322.

\bibitem{kapolka16min}
M.~Kapolka, E.~Pardo, Three-dimensional electromagnetic modeling of practical
  superconductors for power applications, Midterm PhD thesis
  reportArXiv:1605.09610.

\bibitem{Modelling_website}
HTS Modelling Workgroup. \url{http://www.htsmodelling.com}.

\bibitem{kim62PRL}
Y.~B. Kim, C.~F. Hempstead, A.~R. Strnad, Critical persistent currents in hard
  superconductors, Phys. Rev. Lett. 9~(7) (1962) 306--309.

\bibitem{courant}
R.~Courant, D.~Hilbert, Methods of Mathematical Physics, Volume {I},
  Interscience Publishers, New York, 1953.

\bibitem{jackson}
J.~D. Jackson, Classical Electrodynamics, John Wiley {\&} Sons {Inc.}, 3rd
  edition, 1999.

\bibitem{stenvall13IES}
A.~Stenvall, F.~Grilli, M.~Lyly, Current-penetration patterns in twisted
  superconductors in self-field, IEEE Trans. Appl. Supercond. 23~(3) (2013)
  8200105--8200105.

\bibitem{halse70JPD}
M.~R. Halse, {AC} face field losses in a type {II} superconductor, J. Phys. D:
  Appl. Phys. 3 (1970) 717--720.

\bibitem{clem94PRB}
J.~Clem, A.~Sanchez, Hysteretic ac losses and susceptibility of thin
  superconducting disks, Phys. Rev. B 50~(13) (1994) 9355.

\bibitem{Brandt93PRBa}
E.~Brandt, M.~Indenbom, {Type-II-superconductor} strip with current in a
  perpendicular magnetic field, Phys. Rev. B 48~(17) (1993) 12893--12906.

\bibitem{Zeldov94PRB}
E.~Zeldov, J.~R. Clem, M.~{McElfresh}, M.~Darwin, Magnetization and transport
  currents in thin superconducting films, Phys. Rev. B 49~(14) (1994)
  9802--9822.

\bibitem{brandt98PRBa}
E.~H. Brandt, Superconductor disks and cylinders in an axial magnetic field.
  {I}. {Flux} penetration and magnetization curves, Phys. Rev. B 58~(10) (1998)
  6506.

\bibitem{ciftja11PLA}
O.~Ciftja, Coulomb self-energy of a uniformly charged three-dimensional cube,
  Physics Letters A 375~(3) (2011) 766--767.

\end{thebibliography}
\end{document}